\documentclass[a4paper,11pt]{article}
\usepackage{float}
\usepackage{placeins}
\usepackage{jheppub}
\usepackage[normalem]{ulem}
\usepackage{lineno}
\usepackage{siunitx}
\usepackage[caption=false]{subfig}
\usepackage{booktabs}
\usepackage{comment}
\usepackage{adjustbox}
\usepackage{orcidlink}

\usepackage{url} 

\usepackage{amsmath, amssymb, amsthm, amsfonts}

\usepackage{etoolbox}
\usepackage[linesnumbered, ruled, vlined]{algorithm2e}

\usepackage[acronym]{glossaries}
\makeglossaries

\arxivnumber{2608.19064}

\title{\boldmath Exploring the limits of high-energy proton-pion separation in granular calorimeters}

\author[abe]{Andrea De Vita\orcidlink{0009-0004-1790-8629}}
\author[c]{Abhishek\orcidlink{0009-0001-1239-4663}}
\author[de]{Tommaso Dorigo\orcidlink{0000-0002-1659-8727}}
\author[f]{Pietro Vischia\orcidlink{0000-0002-7088-8557}}

\affiliation[a]{CERN, European Organization for Nuclear Research, \\
Esplanade des Particules 1, 1211 Geneva 23, Switzerland}
\affiliation[b]{University of Padova,\\
Department of Physics and Astronomy,\\
Via Marzolo 8, 35131 Padova, Italy}
\affiliation[c] {University of Tennessee, Knoxville, TN, USA}
\affiliation[d]{Lule\aa \, University of Technology,\\
Laboratorievagen 14, 97753 Lule\aa, Sweden}
\affiliation[e]{Istituto Nazionale di Fisica Nucleare,\\
Sezione di Padova, Via Marzolo 8, 35131 Padova, Italy}

\affiliation[f] {Departamento de Física and ICTEA, Universidad de Oviedo,\\
C/ Leopoldo Calvo Sotelo 18, 33007 Oviedo, Spain}

\emailAdd{andrea.de.vita@cern.ch}

\abstract{Highly granular calorimeters give access to detailed information about hadronic-shower development that may enable \gls{pid} beyond their conventional role in energy measurement. We investigate the extent to which this information can distinguish protons from positively charged pions and how the achievable discrimination depends on detector segmentation and particle energy. The study uses \textsc{Geant4} simulations of isolated particles between \SI{10}{GeV} and \SI{100}{GeV} in a homogeneous lead-tungstate calorimeter. A Deep Sets model operating directly on cell positions and detected energy and time is observed to outperform a \gls{bdt} based on reconstructed shower observables.
With \(3\times3\times6~\mathrm{mm}^{3}\) cells, Deep Sets achieves an accuracy of \(93.8\%\) at \SI{10}{GeV}, decreasing to \(67.2\%\) at \SI{100}{GeV}. Shower topology is independently informative, deposited energy provides the largest additional contribution, and timing supplies complementary information. Coarser segmentation reduces discrimination, with performance more sensitive to longitudinal than transverse granularity. These results provide an encouraging benchmark for calorimeter-based hadron identification and motivate its inclusion in the optimization targets of future highly granular calorimeters.
}

\begin{document}
\maketitle
\flushbottom

\newacronym{ml}{ML}{Machine Learning}
\newacronym{ai}{AI}{artificial intelligence}
\newacronym{cnn}{CNN}{Convolutional Neural Network}
\newacronym{gnn}{GNN}{Graph Neural Network}
\newacronym{cgnn}{CGNN}{Convolutional Graph Neural Network}
\newacronym{fcc}{FCC}{Future Circular Collider}
\newacronym{cms}{CMS}{Compact Muon Solenoid}
\newacronym{atlas}{ATLAS}{A Toroidal LHC Apparatus}
\newacronym{dnn}{DNN}{Deep Neural Network}
\newacronym{lhc}{LHC}{Large Hadron Collider}
\newacronym{mip}{MIP}{minimum ionizing particle}
\newacronym{cw}{CW}{Column-Wise ntuple}
\newacronym{cwt}{CWT}{Column-Wise + time smearing ntuple}
\newacronym{hlf}{HLF}{High-Level Feature}
\newacronym{llf}{LLF}{Low-Level Feature}
\newacronym{ecal}{ECAL}{electromagnetic calorimeter}
\newacronym{rw}{RW}{Row-Wise nutple}
\newacronym{sipm}{SiPM}{silicon photon multiplier}
\newacronym{tof}{ToF}{time of flight}
\newacronym{bdt}{BDT}{Boosted Decision Tree}
\newacronym{ei}{EI}{Expected Improvement}
\newacronym{pi}{PI}{Probability of Improvement}
\newacronym{ucb}{UCB}{Upper Confidence Bound}
\newacronym{sma}{SMA}{Simple Moving Average}
\newacronym{bsm}{BSM}{Beyond Standard Model}
\newacronym{sm}{SM}{Standard Model}
\newacronym{eft}{EFT}{Effective Field Theory}
\newacronym{qcd}{QCD}{Quantum Chromodynamics}
\newacronym{sfv}{SFV}{Spontaneous Flavor Violation}
\newacronym{pid}{PID}{particle identification}
\newacronym{cepc}{CEPC}{Circular Electron Positron Collider}
\newacronym{hgcal}{HGCal}{High-Granularity Calorimeter}
\newacronym{lep}{LEP}{Large Electron Positron Collider}
\newacronym{aleph}{ALEPH}{Apparatus for LEP PHysics}
\newacronym{calice}{CALICE}{Calorimeter for Linear Collider Experiment}
\newacronym{pdg}{PDG}{Particle Data Group}
\newacronym{lidar}{LiDAR}{Light Detection and Ranging}
\newacronym{pf}{PF}{Particle Flow}
\newacronym{hep}{HEP}{High Energy Physics}
\newacronym{sota}{SoTA}{State of The Art}

\section{Introduction}

Historically, calorimeters in collider experiments were designed primarily to measure the collective energy of hadronic showers and jets. Over the past two decades, the development of \gls{pf} reconstruction and boosted-object identification has increased the need to resolve the internal structure of particle showers, motivating highly granular calorimeters that operate as imaging detectors. In \gls{pf} algorithms, fine segmentation facilitates the separation of nearby showers and the association of calorimeter deposits with charged-particle tracks, while boosted-object reconstruction benefits from detailed measurements of shower topology. These developments have also encouraged a transition from reconstruction algorithms based on predefined clusters and engineered observables to machine-learning methods operating directly on calorimeter cells or hits. Fine longitudinal and transverse segmentation, potentially combined with precision timing, may therefore enable \gls{pid} directly from calorimetric measurements by exploiting shower shape, energy density, and temporal development in addition to the total deposited energy.

Identifying charged-hadron species using calorimeter information alone nevertheless remains challenging and has received comparatively limited attention~\cite{DeVita2025}. Hadronic showers exhibit large event-by-event fluctuations, while particle-dependent differences in longitudinal development, transverse spread, electromagnetic fraction, local energy density, and timing structure are often subtle. Machine-learning methods are well suited to this task because they can exploit correlations among spatial, temporal, and energy-related observables. Feature-based methods provide an interpretable baseline, whereas models operating directly on calorimeter cells or hits may retain a larger fraction of the available detector information.

In Ref.~\cite{DeVita2025}, we investigated the identification of protons, charged pions, and charged kaons using topological and timing information extracted from highly granular calorimeter primitives. That study demonstrated that calorimeter granularity provides measurable discrimination power and identified physically meaningful observables contributing to charged-hadron separation. However, the dependence of this discrimination power on the incident-particle energy and detector geometry remains insufficiently understood. Characterizing these dependencies is essential both for assessing the applicability of calorimeter-based \gls{pid} and for determining the detector granularity required to preserve the relevant shower information.

In this work, we focus on proton--pion discrimination and evaluate machine-learning classifiers over a range of particle energies and detector geometries. We compare feature-based classifiers with a point-cloud architecture, quantify the effects of longitudinal and transverse segmentation, and investigate the relative contributions of spatial, energy, and timing information. Our objective is to assess the potential of highly granular calorimeters as complementary \gls{pid} systems and to provide guidance for the optimization of future detector designs.

This paper is organized as follows. Section~\ref{sec:related} reviews the relevant developments in calorimeter-based \gls{pid} and machine-learning reconstruction. Sections~\ref{sec:simulation} and~\ref{sec:ml_strategy} describe the detector simulation, data representations, and machine-learning strategy. The results are presented in Section~\ref{sec:results}, discussed in Section~\ref{sec:discussion}, and summarized in Section~\ref{sec:conclusions}.

\section{Related Work}
\label{sec:related}

This section reviews developments relevant to particle reconstruction and identification
with highly granular calorimeters. We begin with conventional \gls{pid} techniques and physics-
motivated methods for exploiting calorimeter shower properties. We then discuss the
progression from classifiers based on engineered observables to machine-learning models
operating directly on cell-level measurements. Broader developments in calorimeter re-
construction are included where they motivate detector representations and architectures
relevant to \gls{pid}.

\subsection{Calorimeter Observables For Particle Identification}

Traditional \gls{pid} in \gls{hep} relies primarily on dedicated detector systems and observables such as ionization energy loss~\cite{doi:10.1142/S0217751X26300115}, time-of-flight measurements~\cite{KLEMPT1999542}, Cherenkov radiation~\cite{PAPANESTIS2020162004}, and transition radiation~\cite{ANDRONIC2012130}. Calorimeters provide complementary information through particle-dependent differences in shower development and detector response.

Electromagnetic showers are generally compact, whereas hadronic showers tend to be broader, develop more deeply, and exhibit larger event-by-event fluctuations and an invisible-energy component. Muons typically produce a minimum-ionizing signature, occasionally accompanied by localized radiative losses. These characteristic responses motivate observables such as longitudinal energy fractions, lateral shower widths, shower depth, energy leakage, cluster isolation, and, for charged particles, the ratio of calorimeter energy to track momentum~\cite{RevModPhys.75.1243}. In experimental applications, calorimeter observables are commonly combined with information from other detector subsystems. For example, \acrshort{atlas} electron identification uses shower-shape and hadronic-leakage variables together with tracking, track--cluster matching, and transition-radiation measurements~\cite{Aad:1694142}.

Calorimeter-reconstruction algorithms also exploit the internal structure of particle showers. Three-dimensional topological clustering groups cells according to signal significance and spatial connectivity, suppressing noise while retaining the principal shower features~\cite{aad2017topocluster}. Local software compensation instead reweights energy deposits according to their local density to improve the energy resolution of non-compensating calorimeters~\cite{Adloff2012}. These methods demonstrate the reconstruction value of shower topology and energy-density information, but rely on predefined observables, explicit clustering rules, and detector-specific optimization. Highly granular calorimeters provide richer measurements of the same properties, motivating approaches that learn directly from the cell-level detector response.

\subsection{Machine Learning With Highly Granular Calorimeters}

Machine-learning applications to calorimeter data have progressed from likelihood discriminants and \glspl{bdt} based on engineered shower observables to deep-learning models operating directly on cells or hits. These models can learn spatial features, particle identities, and energy corrections from low-level detector measurements~\cite{krause_2026_21626667}. For regularly segmented calorimeters, convolutional neural networks provide a natural representation by treating the cell-level response as an image. Reference~\cite{Belayneh_2020}, for example, demonstrated simultaneous particle classification and energy regression from calorimeter shower images. Fixed-grid representations, however, are less natural for sparse responses, irregular detector geometries, or systems containing regions with different granularities.

Set-, point-cloud-, and graph-based architectures address these limitations by representing a shower as an unordered collection of detector measurements. Such models can learn correlations among cell positions, energies, and times while preserving the sparse structure of highly granular calorimeter data. GravNet-style message passing combined with object condensation, for example, has been used to cluster overlapping showers and jointly predict their identities and properties~\cite{qasim2022multiparticle,kieseler2020object}. Although developed for broader reconstruction tasks, these approaches demonstrate the suitability of point-based representations for extracting information directly from granular detector measurements. Their application nevertheless requires careful consideration of simulation fidelity, calibration, computational cost, and robustness outside the training phase space.

Highly granular calorimeters have also enabled inference tasks beyond conventional energy reconstruction. Deep-learning methods have shown that the spatial distribution of radiative losses can be used to estimate the energy of multi-TeV muons with a precision exceeding that of track-curvature measurements at the highest energies~\cite{mueloss1}; comparable results have been obtained using a (k)-nearest-neighbor approach~\cite{mueloss2}. These studies demonstrate that fine-grained shower topology contains information beyond the total deposited energy.

\section{Simulation And Dataset Preparation}
\label{sec:simulation}

In order to study the physical processes occurring inside the calorimeter, a detailed simulation framework is developed using \texttt{GEANT4}.

The aim is to characterize the behavior of the showers initiated by hadrons. The simulation does not include the effects of detector readout, such as electronic noise or signal shaping. These factors will be considered in future studies.

\subsection{Simulation Setup}

The \texttt{GEANT4} detector simulation uses the \texttt{FTFP\_BERT} physics list, following the approach described in Ref.~\cite{DeVita2025}. Primary protons (\(p\)) and charged pions (\(\pi^+\)) are generated with total energies of \SI{100}{\GeV}, \SI{50}{\GeV}, \SI{25}{\GeV}, and \SI{10}{\GeV}. The particles originate \SI{3}{\meter} upstream of the calorimeter and propagate in vacuum parallel to the \(z\)-axis without angular spread. Each event only includes one generated particle.

The calorimeter is modeled as a homogeneous lead tungstate (\(\mathrm{PbWO_4}\)) detector segmented into \(100 \times 100 \times 200\) cells, each measuring \(3 \times 3 \times 6\,\mathrm{mm}^3\). The resulting detector dimensions are \(300 \times 300 \times 1200\,\mathrm{mm}^3\), corresponding to a lateral extent of \(7.66\) Molière radii and a longitudinal depth of \(5.92\) nuclear interaction lengths. Shower containment and its implications are discussed in Ref.~\cite{DeVita2024Thesis}.

\begin{figure}[ht]
    \centering
    \includegraphics[width=0.95\textwidth]{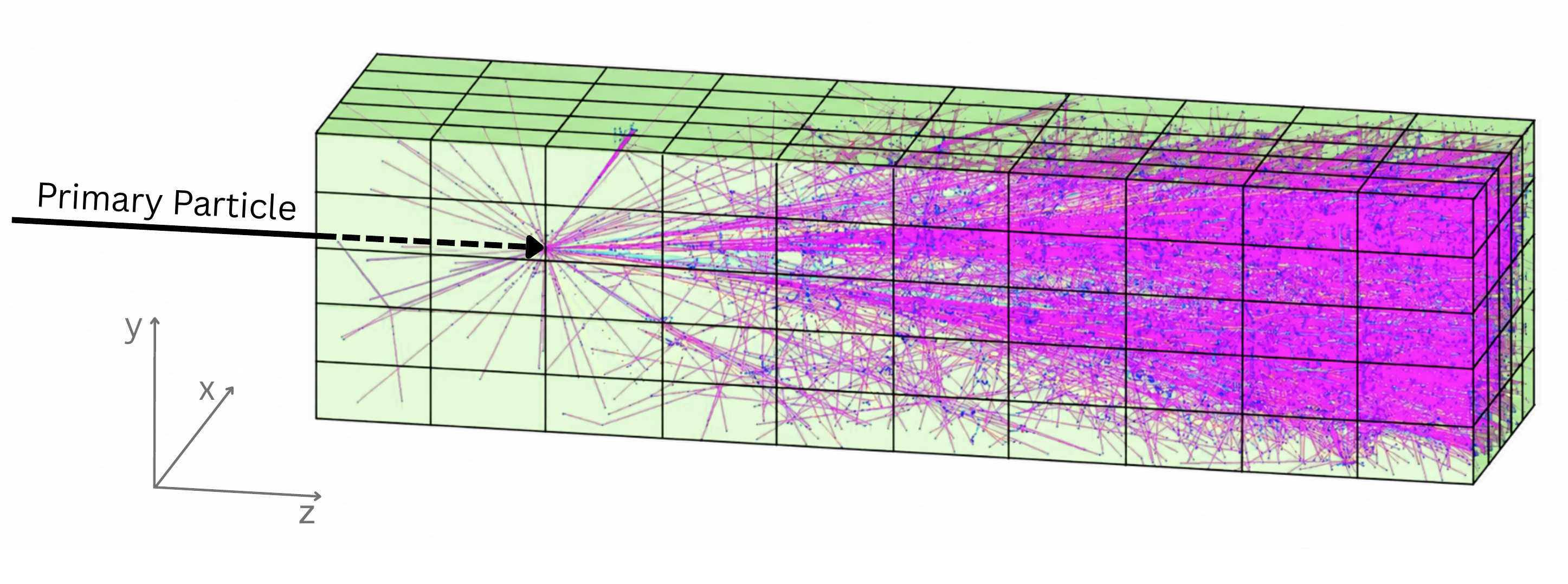}
    \caption{\texttt{GEANT4} event display of a simplified calorimeter used in this study. The subdivision of the calorimeter volume in cells is only indicative, showing much coarser cells than those handled by the simulation~\cite{DeVita2024Thesis}.}
    \label{fig:eventDisplay}
\end{figure}

For each simulation step, \texttt{GEANT4} records the particle identity (\acrshort{pdg} code), momentum, kinetic energy loss, deposited energy, global time, and cell coordinates. In this work, only the quantities accessible in realistic detector conditions are used for the analysis, namely the deposited energy, the hit position, and the global time (see Fig.~\ref{fig:eventDisplay} for an example of event display).

The simulation framework was validated by verifying energy conservation and the consistency of the initial particle momentum with relativistic kinematics. The overall setup closely follows the methodology of Ref.~\cite{DeVita2025}.

\subsection{Data Preprocessing}
\label{sec:preprocessing}

The simulation output is preprocessed to reduce storage requirements and produce input formats suitable for the machine-learning approaches adopted in this work. The preprocessing pipeline follows the procedure described in Ref.~\cite{DeVita2024Thesis}.

To suppress low-energy deposits and reduce the output size, only simulation steps satisfying \(\texttt{TotalEnergyDeposit} \geq \SI{1}{\keV}\) or \(\texttt{DeltaKineticEnergy} \geq \SI{1}{\keV}\) are retained. The selected events are then converted from the original row-wise representation to a compact column-wise format, improving both storage efficiency and I/O performance.

Starting from this dataset, two parallel processing branches are constructed. In the first, the recorded hit times are smeared according to a Gaussian distribution with a resolution of \(\sigma=\SI{30}{\pico\second}\), consistent with current detector technologies~\cite{PerezLara2024}. The resulting column-wise files are used to compute the \glspl{hlf} introduced in Ref.~\cite{DeVita2025}.

In the second branch, no time smearing is applied. For each event, the \glspl{llf} associated with all active calorimeter cells are stored in a \texttt{pandas.DataFrame}. Each cell is characterized by its spatial position, deposited energy, and characteristic time computed from the original simulated hit times. These data are subsequently interpreted as enriched point clouds and used as input to the point-cloud models described in Sec.~\ref{sec:ml_strategy}.

\begin{figure}[ht]
    \centering
    \includegraphics[width=0.9\textwidth]
    {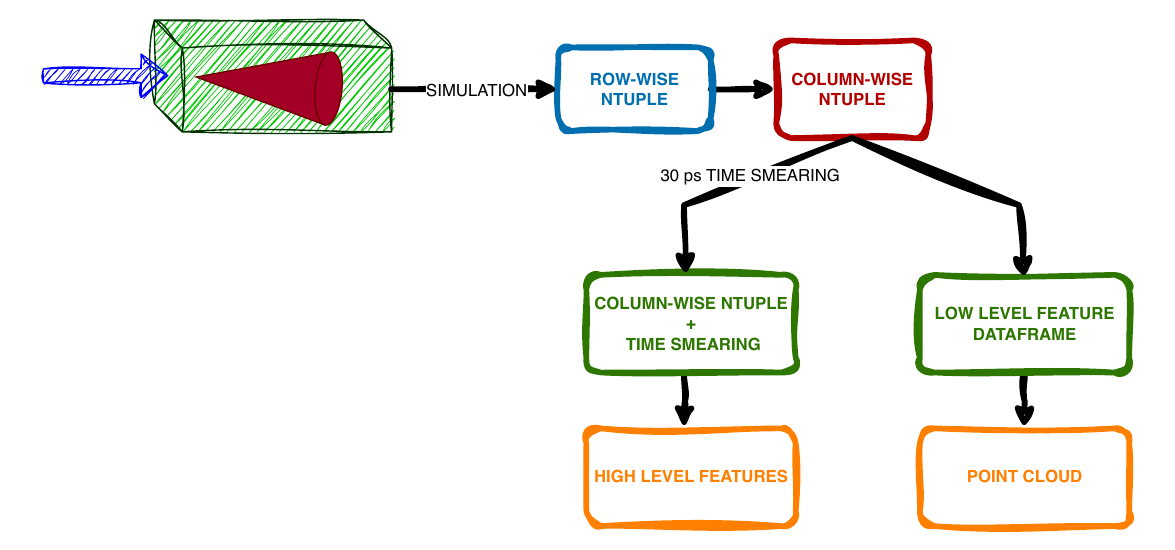}
    \caption{Overview of the data-reduction and preprocessing pipeline~\cite{DeVita2024Thesis}.}
    \label{fig:preprocessingPipeline}
\end{figure}

The preprocessing workflow is summarized in Fig.~\ref{fig:preprocessingPipeline}. Different detector granularities are obtained after simulation by merging adjacent cells. Since all segmentation schemes are derived from the same fully granular dataset, the corresponding samples are intrinsically correlated, as discussed in Ref.~\cite{DeVita2024Thesis}.

\subsection{Shower Representation}
\label{sec:shower_representation}

Calorimeter showers are represented using two complementary approaches, following the methodology introduced in Ref.~\cite{DeVita2024Thesis}: a vector of \glspl{hlf} and an enriched point cloud that preserves the detector granularity.

Both representations are constructed from the same cell-level observables, although the treatment of timing information differs between the two processing branches. The position of each active cell is identified with the coordinates of its geometrical center. The deposited energy is obtained by summing all energy deposits recorded within the cell:
\begin{equation}
    E_i^{\mathrm{cell}}
    =
    \sum_j E_{ij},
\end{equation}
where \(E_{ij}\) denotes the \(j\)-th energy deposition in cell \(i\). In this study, the measured cell energy is identified with the simulated deposited energy, corresponding to a unit conversion factor between the two quantities.

The characteristic time of cell \(i\) is defined as the energy-weighted average of the corresponding hit times:
\begin{equation}
    t_i^{\mathrm{cell}}
    =
    \frac{\sum_j t_{ij} E_{ij}}
         {\sum_j E_{ij}},
\end{equation}
where \(t_{ij}\) is the time associated with the \(j\)-th energy deposition in the cell. For the \gls{hlf} representation, the hit times are smeared with a Gaussian resolution of \(\sigma=\SI{30}{\pico\second}\), whereas the point-cloud representation uses the original unsmeared simulation times.

\begin{figure*}[ht]
    \centering

    \adjustbox{valign=c}{
        \includegraphics[height=5cm]
        {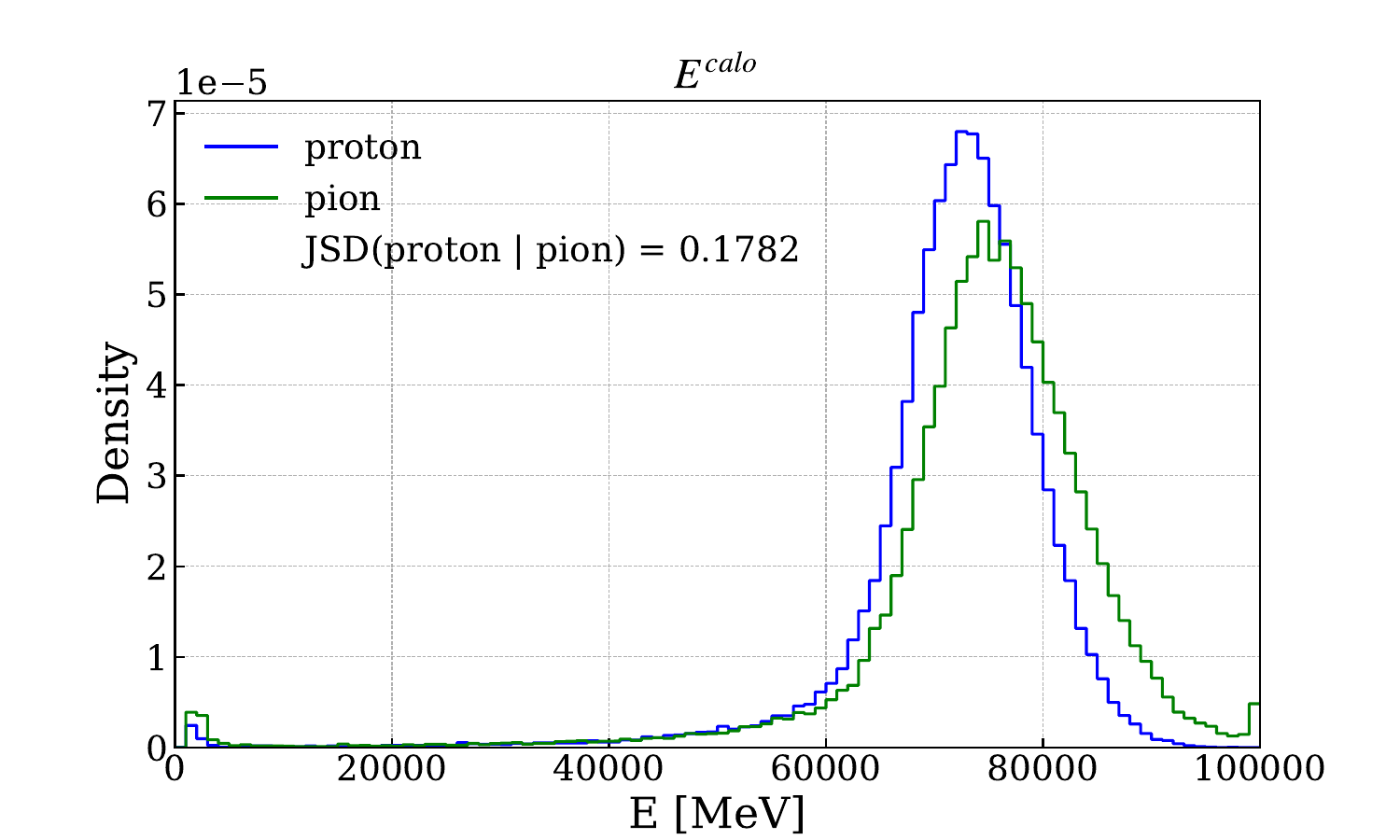}
    }
    \hfill
    \adjustbox{valign=c}{
        \includegraphics[height=5cm]
        {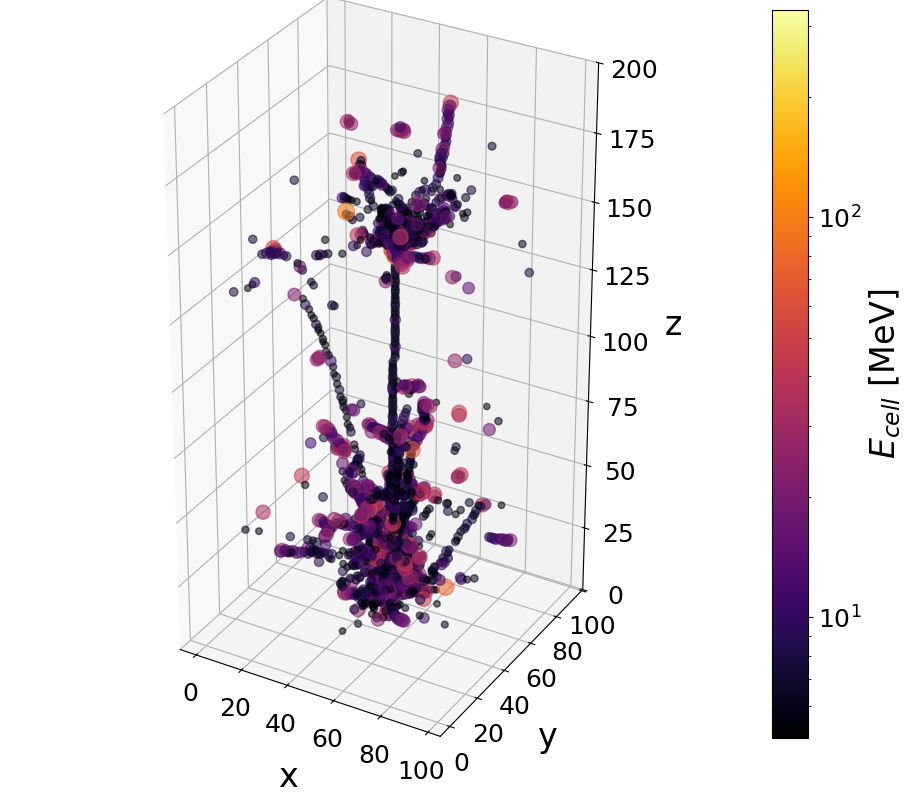}
    }

    \caption{
        (\textbf{Left}) Distribution of the total energy deposited in the calorimeter for proton- and pion-induced showers at \SI{100}{\GeV}, using a cell size of \(3 \times 3 \times 12~\mathrm{mm}^{3}\). The corresponding Jensen--Shannon divergence quantifies the similarity between the two distributions~\cite{Lin1991,DeVita2025}.
        (\textbf{Right}) Example of the enriched point-cloud representation of a proton-induced shower at \SI{100}{\GeV}~\cite{DeVita2024Thesis}.
    }
    \label{fig:event_sparsity}
\end{figure*}

These observables provide a common detector-level description of the calorimeter response from which the two shower representations are derived. The \gls{hlf} representation summarizes each shower through observables describing its global properties, quantities related to the reconstructed first nuclear interaction, the local energy distribution, and its longitudinal and transverse development. An example is shown in Fig.~\ref{fig:event_sparsity} on the left, while the complete feature set is defined in Ref.~\cite{DeVita2025}.

Alternatively, each shower is represented as an enriched point cloud,
\begin{equation}
    \mathcal{S}
    =
    \left\{
        \left(
            x_i,
            y_i,
            z_i,
            E_i^{\mathrm{cell}},
            t_i^{\mathrm{cell}}
        \right)
    \right\}_{i=1}^{N_{\mathrm{cell}}},
\end{equation}
where each active calorimeter cell corresponds to one point. The coordinates \((x_i,y_i,z_i)\) encode the detector geometry, while \(E_i^{\mathrm{cell}}\) and \(t_i^{\mathrm{cell}}\) represent the deposited energy and characteristic time, respectively. This representation preserves the calorimeter granularity and is naturally suited to permutation-invariant architectures such as Deep Sets~\cite{DeVita2024Thesis}. An example is shown in Fig.~\ref{fig:event_sparsity} on the right.

\section{Machine Learning Strategy}
\label{sec:ml_strategy}

The choice of the machine learning model depends on the adopted shower representation. In this study, \glspl{hlf} are analyzed using XGBoost, while the point-cloud representation is processed with a Deep Sets architecture, which is specifically designed for unordered sets of variable cardinality. The implementation follows the methodology developed in Ref.~\cite{DeVita2024Thesis}. While the chosen models may not achieve the absolute highest performance in the classification task which more tailored models and larger analyzed datasets could produce, we believe that their performance is sufficient to approximate that ceiling, such that useful considerations can be made concerning their value.

\subsection{Machine Learning Models}

The \gls{hlf} representation is classified using a \gls{bdt} implemented with the XGBoost library~\cite{Chen2016}. Gradient boosting constructs an ensemble of shallow decision trees, where each successive tree is trained to correct the residual errors of the current model~\cite{Friedman2001}. Besides providing competitive classification performance, XGBoost offers feature-importance metrics that allow the physical relevance of the engineered observables to be investigated.

The point-cloud representation is processed with a Deep Sets architecture~\cite{zaheer2018}, which satisfies permutation invariance by construction and is therefore well suited to calorimeter showers represented as unordered sets of active cells. Each point is described by its position, deposited energy, and characteristic time, while a permutation-invariant pooling operation produces the global event representation used for classification. Additional implementation details are reported in Ref.~\cite{DeVita2024Thesis}.

\subsection{Hyperparameter Optimization}

The model hyperparameters were optimized before the final training. For XGBoost, an exhaustive grid search with three-fold cross-validation was performed over the main tree and regularization parameters. For Deep Sets, Bayesian optimization was employed to determine the optimal learning rate and batch size while keeping the network architecture fixed. The resulting hyperparameter configurations are reported in Tables~\ref{tab:xgboost_hyperparameters} and~\ref{tab:deepsets_hyperparameters}. Further details of the optimization procedure can be found in Ref.~\cite{DeVita2024Thesis}.

\begin{table}[ht]
\centering

\begin{minipage}[t]{0.47\textwidth}
\centering
\caption{Hyperparameter ranges adopted for the optimized XGBoost models.}
\vspace{4mm}
\label{tab:xgboost_hyperparameters}
\begin{tabular}{lc}
\toprule
\textbf{Hyperparameter} & \textbf{Value} \\
\midrule
Learning rate & $10^{-3}$ \\
Max depth & $7$--$10$ \\
Trees & $3500$--$4500$ \\
Column sampling & $0.6$ \\
L1 regularization & $0.01$--$0.1$ \\
L2 regularization & $2$ \\
\bottomrule
\end{tabular}
\end{minipage}
\hfill
\begin{minipage}[t]{0.50\textwidth}
\centering
\caption{Final Deep Sets architecture and training hyperparameters.}
\vspace{4mm}
\label{tab:deepsets_hyperparameters}
\begin{tabular}{lc}
\toprule
\textbf{Parameter} & \textbf{Value} \\
\midrule
Input features & $(x,y,z,E,t)$ \\
Hidden layers & $[126,168,43,84,22]$ \\
Output neurons & $2$ \\
Pooling & Mean \\
Optimizer & AdamW \\
Batch size & $86$ \\
Learning rate & $2.43\times10^{-3}$ \\
Loss function & CrossEntropyLoss \\
Activation & LeakyReLU \\
\bottomrule
\end{tabular}
\end{minipage}

\end{table}

\subsection{Performance Metrics}

Model performance is evaluated using several complementary classification metrics. Confusion matrices provide a direct summary of correctly and incorrectly classified events, while ROC curves assess the discrimination power of the models over the full range of classification thresholds. For the XGBoost classifier, the relative contribution of each input variable is quantified through the feature importance based on the gain metric. Since no equivalent intrinsic ranking exists for Deep Sets, its feature relevance is instead investigated through dedicated ablation studies (see Sec.~\ref{sec:singleFeatureAnalysis}). In addition, the trade-off between classification purity and event selection is characterized by accuracy--efficiency curves obtained by varying the decision threshold on the predicted class probabilities. Statistical uncertainties on efficiencies, accuracies, and ROC curves are estimated assuming binomial statistics. In particular, confidence intervals for the accuracy are computed using the Clopper--Pearson method, following the procedure described in Ref.~\cite{DeVita2024Thesis}.

~
\section{Results}
\label{sec:results}

This chapter presents the results of the two complementary classification studies introduced in Sec.~\ref{sec:ml_strategy}. The first establishes a baseline using the set of \glspl{hlf} defined in Ref.~\cite{DeVita2025}, together with a \gls{bdt} classifier, to quantify the impact of calorimeter segmentation on proton--pion separation at \SI{100}{GeV}. The second extends the analysis to a point-cloud representation using a Deep Sets architecture, enabling a more expressive description of shower topology over a broader energy range.

\subsection{Baseline Analysis With High-Level Features At \SI{100}{GeV}}
\label{sec:resultsHLF}

This section establishes the baseline performance using \glspl{hlf} extracted from calorimeter simulations at \SI{100}{GeV}. Classification is performed with a \gls{bdt}, which combines competitive discrimination performance with straightforward interpretation through feature importance rankings.

The training and validation datasets contain 80,000 and 20,000 events, respectively, with each dataset containing equal proportions of protons and pions.
The finest segmentation considered is $\{100,100,100\}$, corresponding to a minimum cell size of $3\times3\times12~\mathrm{mm}^3$; coarser detector geometries are obtained by progressively merging neighboring cells.

\paragraph{Feature Importance.}

\begin{figure}[b]
    \centering
    \includegraphics[width=0.72
    \textwidth]{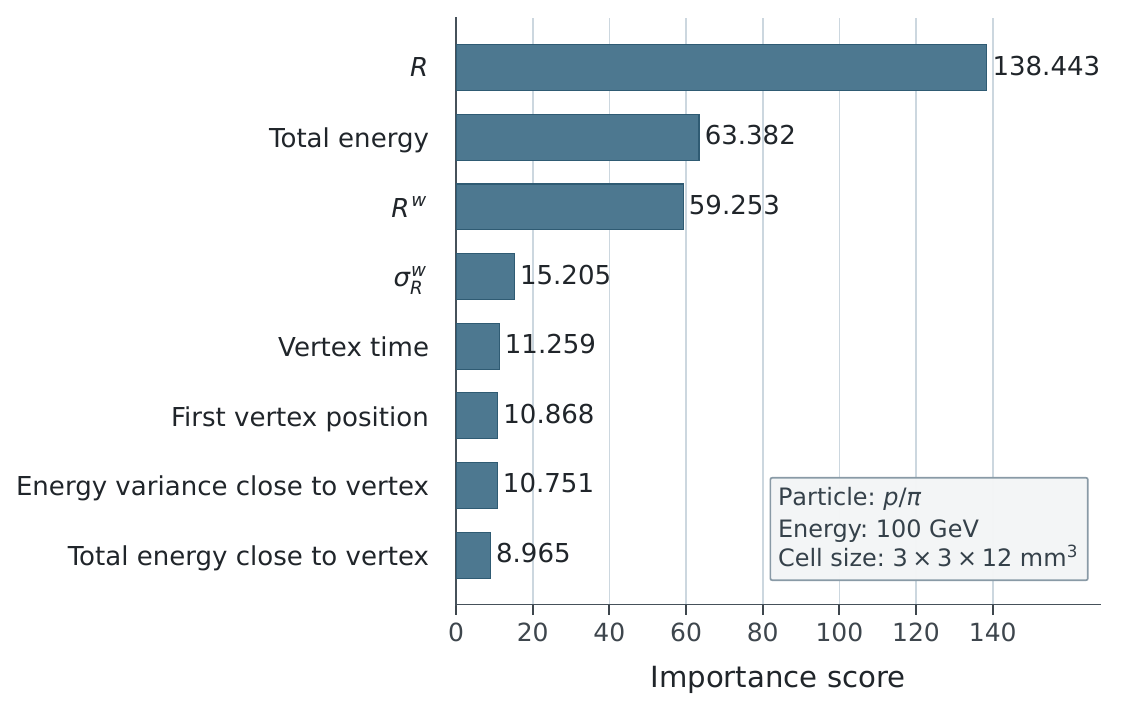}
    \caption{Feature importance ranking for the \gls{bdt} classifier at the finest segmentation, computed using the binary tree gain metric.}
    \label{fig:featureImportanceBaseline}
\end{figure}

Figure~\ref{fig:featureImportanceBaseline} shows the feature-importance ranking obtained at the finest segmentation. The total deposited energy is among the most discriminating observables. As shown in Fig.~\ref{fig:event_sparsity} on the left, pion showers deposit, on average, more visible energy than proton showers. This difference is related to the electromagnetic fraction, $f_{\mathrm{em}}$, of the shower~\cite{Akchurin1998}. In proton-induced showers, baryon-number conservation requires a baryon to remain in the final state, suppressing processes in which the incident proton transfers a large fraction of its energy to $\pi^0$ production~\cite{Akchurin1998}. Consequently, events with very large $f_{\mathrm{em}}$ are less probable for protons. Pions carry no baryon number and can instead transfer a large fraction of their energy to one or more $\pi^0$ mesons, for example through the charge-exchange process $\pi^- + p \rightarrow \pi^0 + n$. The subsequent decay $\pi^0 \rightarrow \gamma\gamma$ feeds the electromagnetic component, producing a larger calorimeter response and accounting for the higher-energy part of the pion distribution~\cite{DeVita2025,Akchurin1998}.

Several observables describing the electromagnetic component of the shower are also ranked among the most important features. In particular, the shower radius $R$, its energy-weighted counterpart $R^w$, and the corresponding standard deviation $\sigma_R^w$ all appear among the top-ranked variables~\cite{DeVita2025}. This behavior is consistent with the underlying shower physics: the electromagnetic component, driven primarily by $\pi^0$ production, produces a more collimated energy deposition. The larger electromagnetic fraction of pion showers therefore results in systematically smaller shower radii than those observed for protons~\cite{DeVita2025}.

Four \glspl{hlf} associated with the first nuclear interaction vertex, the location of the earliest strong interaction of the primary incident particle with a nucleus of the calorimeter material, are adopted from Ref.~\cite{DeVita2025}: the reconstructed vertex position, the deposited energy around the vertex, the vertex time, and the energy variance in that region. Their comparatively lower importance likely reflects two compounding effects. First, these vertex-related observables~\cite{DeVita2025} are expected to benefit from finer longitudinal granularity than that considered here. Second, the vertex-finding algorithm implemented in this study is not perfectly accurate (see Ref.~\cite{DeVita2024Thesis}), reducing the discriminating power of the corresponding vertex-related features~\cite{DeVita2025}, which depend on the precise localization of the first interaction.

\paragraph{Model Performance And Dependence On Segmentation.}

For each event, the classifier assigns a probability to each particle hypothesis. The larger of the two probabilities, referred to as the \emph{winning probability}, determines both the predicted class and the confidence of the prediction.

\begin{figure*}[t]
    \centering

    \includegraphics[width=0.8\textwidth]{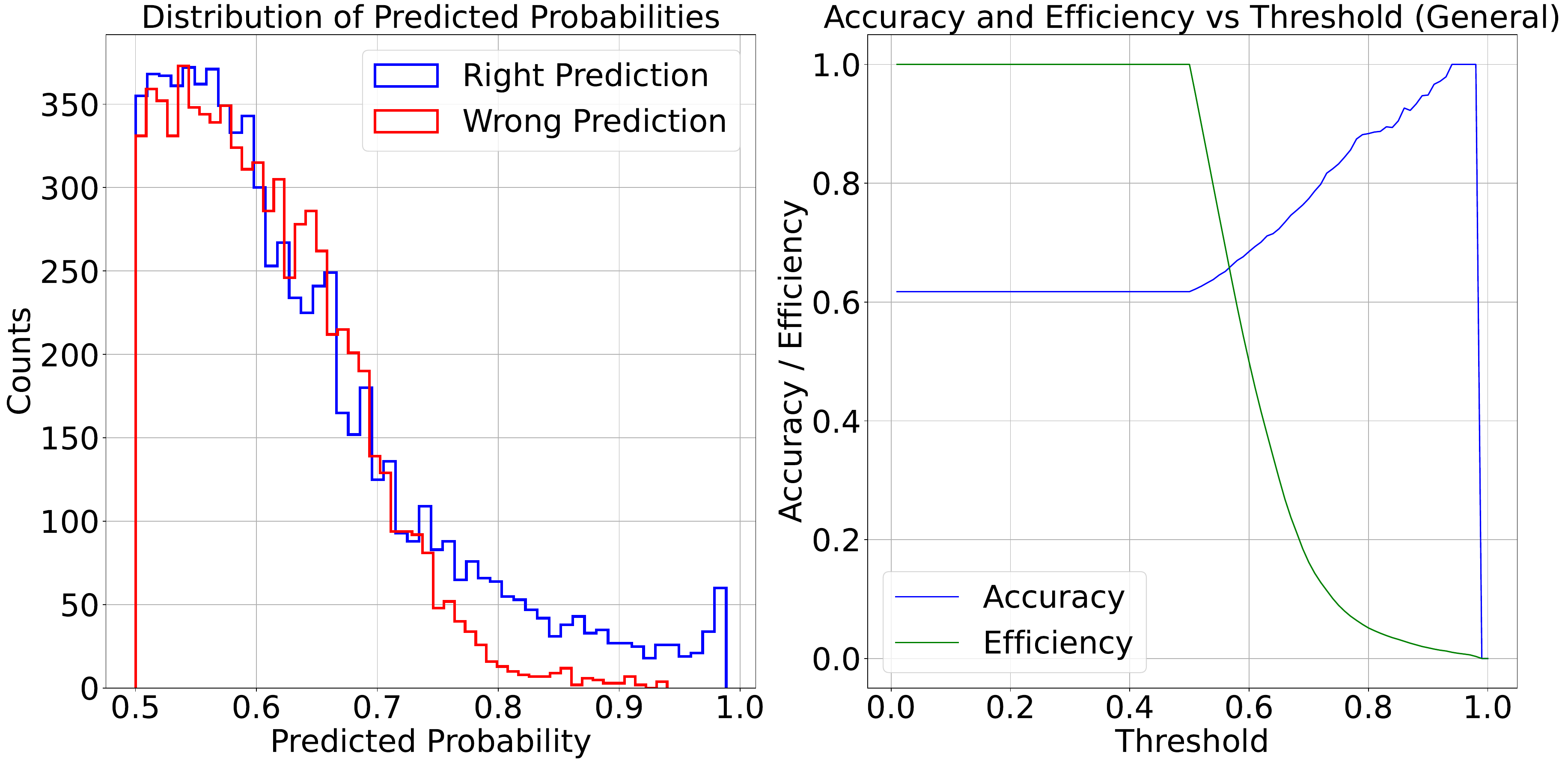}

    \vspace{4mm}

    \includegraphics[width=\textwidth]{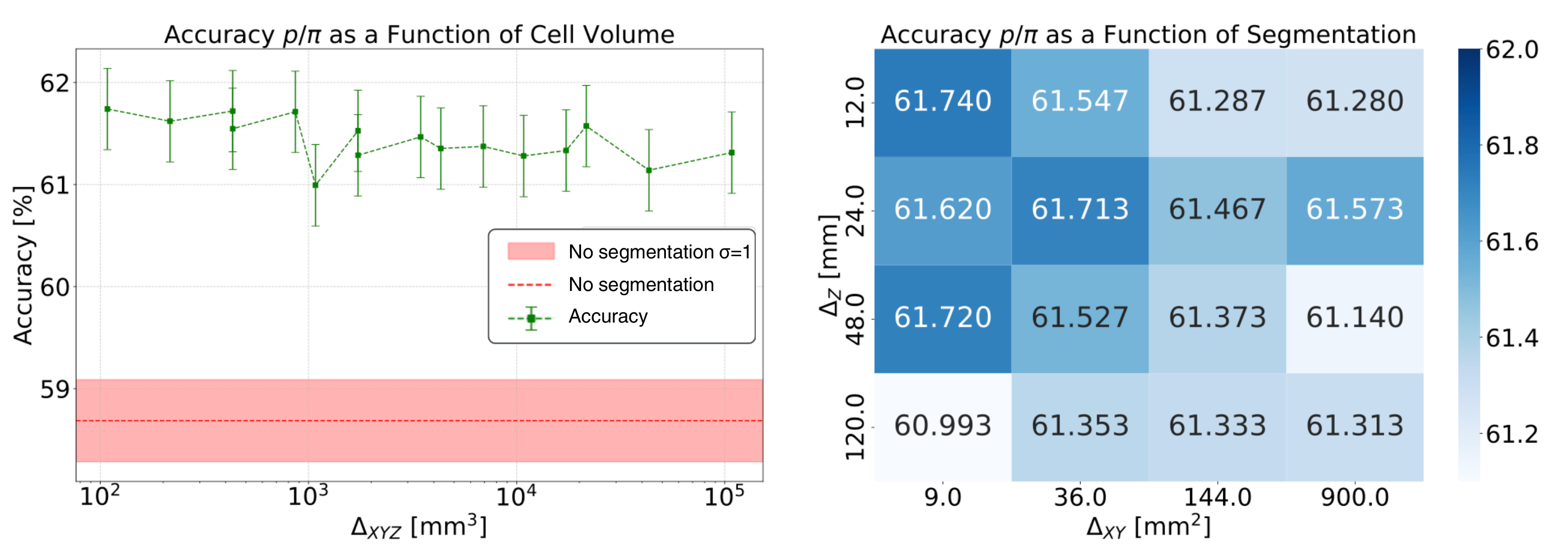}

    \caption{Baseline performance of the \gls{bdt} classifier. (\textbf{Top}) Distribution of the winning probability together with the corresponding accuracy--efficiency curves. (\textbf{Bottom}) Dependence of the classification accuracy on calorimeter segmentation, shown as a function of total cell volume, and the complete scan of segmentation configurations.
    }
    \label{fig:baselinePerformance}
\end{figure*}

The upper panels of Fig.~\ref{fig:baselinePerformance} summarize the baseline classifier behavior. The distribution of winning probabilities shows that correctly classified events cluster near unity, whereas misclassified events are associated with lower confidence values. Applying a threshold on the winning probability therefore improves the classification purity at the expense of signal efficiency. The corresponding accuracy--efficiency curves show that the proton efficiency decreases more rapidly than the pion efficiency, indicating that pion candidates are generally identified with higher confidence.

One of the primary goals of this study is to quantify how calorimeter granularity influences the achievable particle-identification performance. Introducing calorimeter segmentation increases the classification accuracy from the homogeneous baseline of 58.7\% to approximately 61.4\%. Figure~\ref{fig:baselinePerformance} shows that this improvement gradually decreases as either the transverse or longitudinal cell dimensions become larger, while the dependence on the total cell volume exhibits an almost monotonic degradation.

As discussed in Ref.~\cite{DeVita2025}, measurements obtained for different segmentation configurations are strongly correlated because they originate from the same underlying events. Consequently, even relatively small differences between neighboring points represent genuine effects of detector granularity rather than independent statistical fluctuations. The accompanying heatmap provides a compact overview of the performance achieved across all tested segmentation configurations.

\subsection{Analysis With Point Clouds As A Function Of Energy}
\label{sec:resultsPointCloud}
% ─────────────────────────────────────────────────────────────────────────────

The second analysis replaces the hand-crafted \glspl{hlf} with a point cloud representation of the calorimeter shower (described in Sec.~\ref{sec:shower_representation}), processed by a Deep Sets architecture. This approach makes no prior assumptions about the shower structure and is therefore capable of exploiting topological information that may be inaccessible to aggregate features. The study covers four energy points, 10, 25, 50, and \SI{100}{GeV}, and characterizes both the impact of calorimeter granularity and the evolution of classification performance with energy.

The dataset is larger than the one used in Sec.~\ref{sec:resultsHLF}: 100\,000 events per particle type, split 80/20 into training and validation sets. The finest segmentation ($\{100,100,200\}$) is equivalent cells in $x$ and $y$ as before, but twice as fine in $z$ (minimum cell size $3\times3\times6\;\text{mm}^3$). Coarser segmentations are again obtained by merging adjacent cells. All models are trained for 60 epochs.

\subsubsection{Event Pre-Processing: Energy Threshold On Cell Selection}
\label{sec:energyThreshold}

\begin{figure}[t]
    \vspace{-30pt}
    \centering
    \subfloat{
        \includegraphics[width=0.46\textwidth]{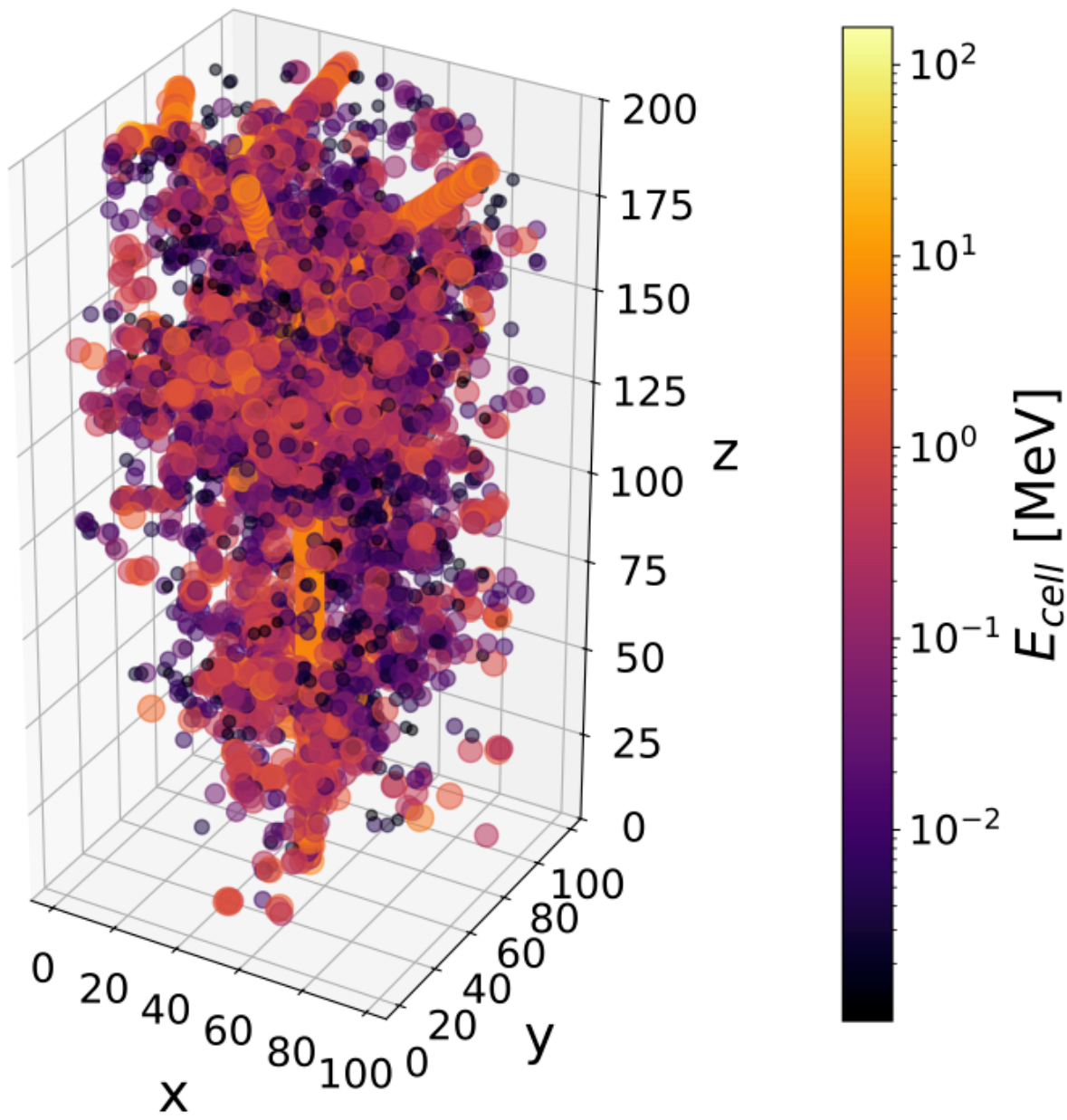}
    }
    \subfloat{
        \includegraphics[width=0.46\textwidth]{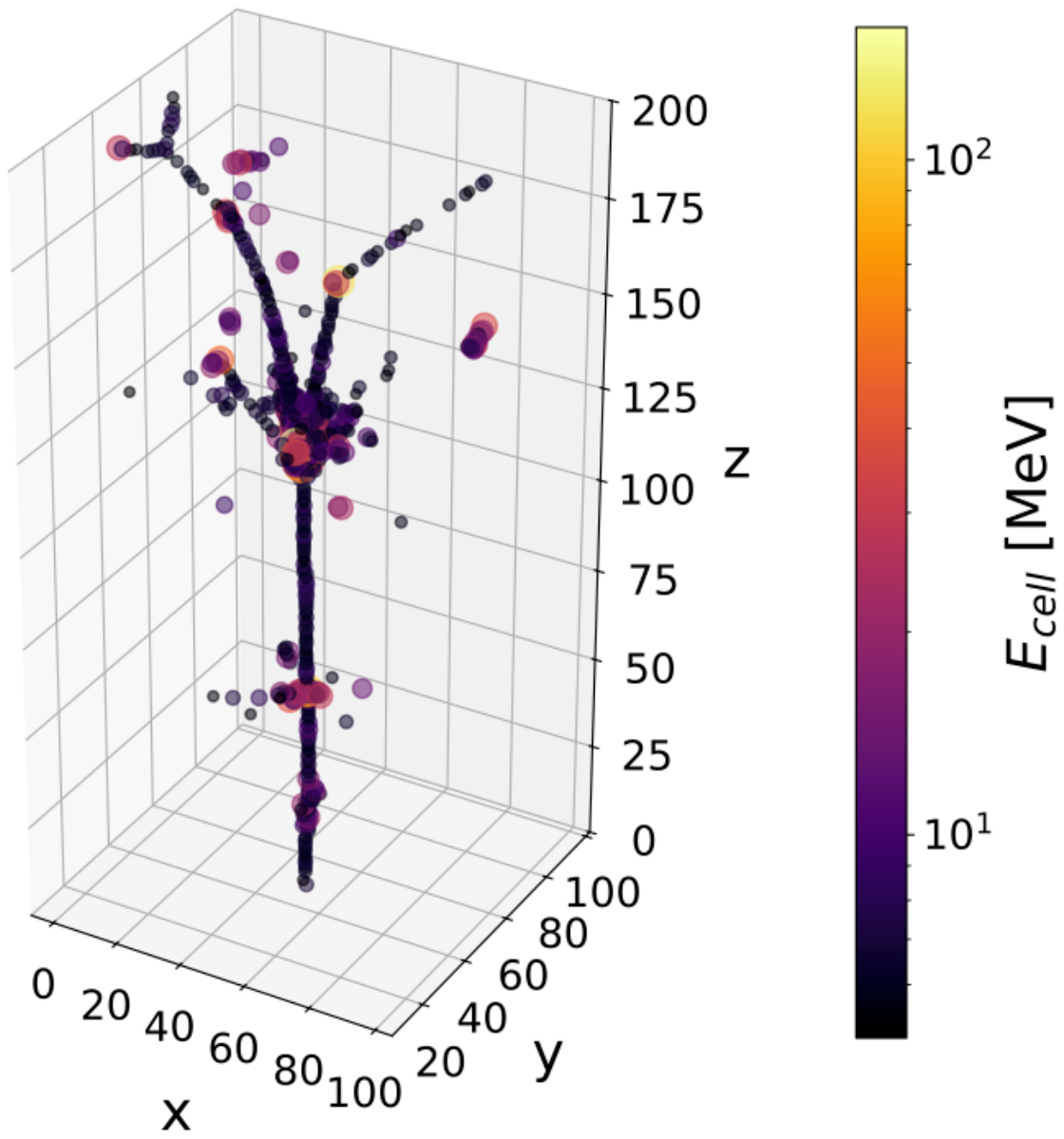}
    }
    \vspace{-10pt}
    \caption{%
        Point cloud representations of the same $\pi^+$ shower at \SI{25}{GeV}.
        Point size is proportional to $E_\text{cell}$; cell xyz coordinates refer
        to cell index. \textbf{(Left)} No energy threshold applied; the cloud is
        dominated by low-energy cells. \textbf{(Right)} Threshold of
        $E_\text{cell} > \SI{5}{MeV}$ applied; the shower topology is
        substantially cleaner.
    }
    \label{fig:thr_effect}
\end{figure}

Prior to training, each event is pre-processed by discarding calorimeter cells whose deposited energy falls below a threshold $E_\text{cell} > \SI{5}{MeV}$. This step serves two purposes. First, it reduces the contribution of low-energy cells that do not carry significant physics information, making the topological structure of the shower more evident. As illustrated in Fig.~\ref{fig:thr_effect}, without any threshold the point cloud is dominated by diffuse, low-energy deposits that largely obscure the shower core; with the threshold applied, the principal features of the shower become clearly visible. Second, discarding near-empty cells increases the sparsity of the data representation, substantially reducing storage and processing requirements.

\begin{figure}[ht]
    \centering
    \includegraphics[width=0.55\textwidth]{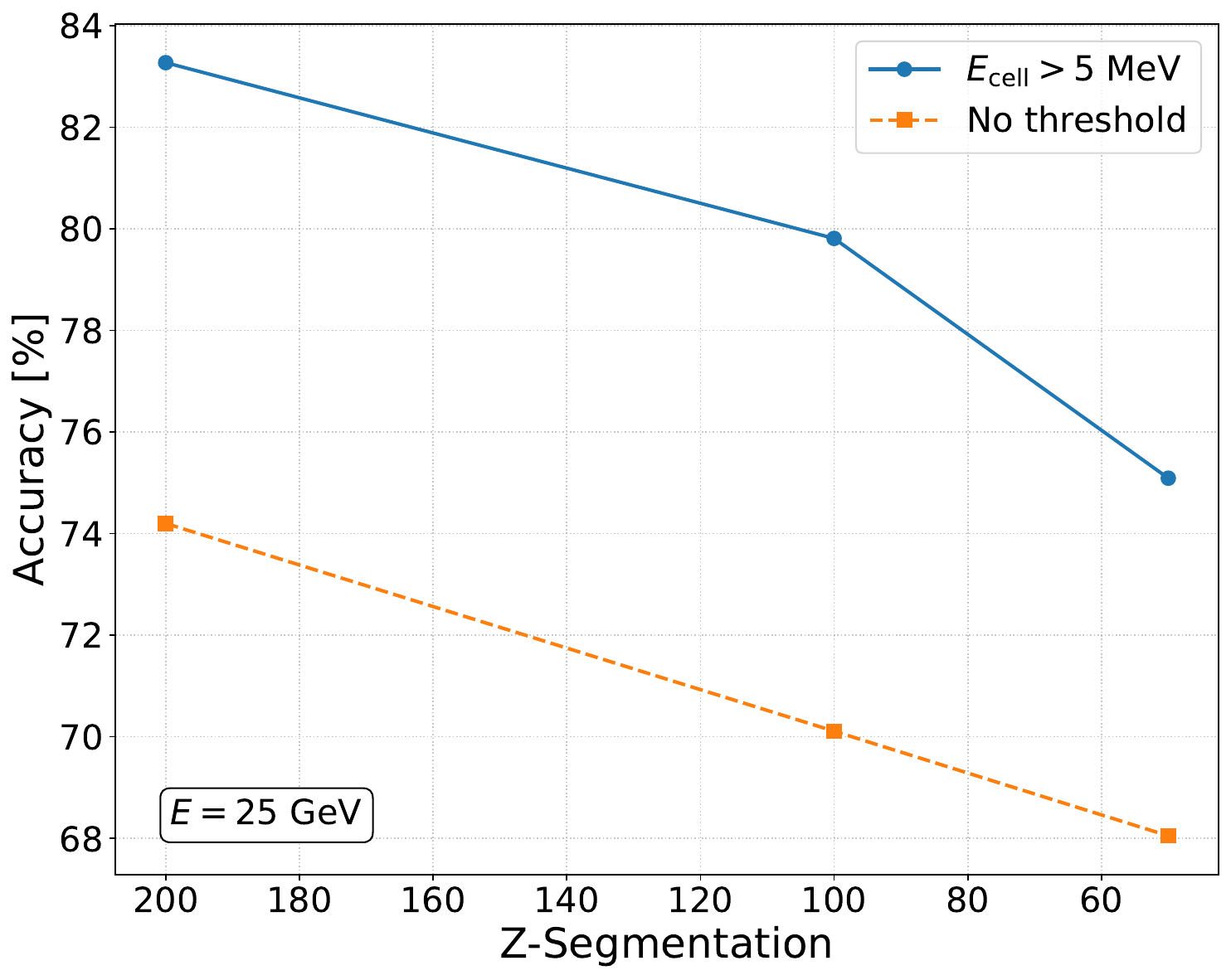}
    \caption{%
        Accuracy as a function of longitudinal cell size for $p/\pi$ classification at \SI{25}{GeV}, with and without the cell energy threshold $E_\text{cell} > \SI{5}{MeV}$.
    }
    \label{fig:threshold_training}
\end{figure}

The quantitative impact of the threshold on model performance is shown in Fig.~\ref{fig:threshold_training} for the $p/\pi$ task at \SI{25}{GeV}. The improvement in accuracy is non-negligible across all longitudinal segmentations, confirming that the threshold genuinely enhances physics-relevant information rather than merely reducing noise. At the same time, the performance gap observed without the threshold suggests that the model is sensitive to low-energy contributions and is not inherently robust to this form of noise. A larger training dataset could potentially mitigate this sensitivity; all subsequent results implicitly assume the \SI{5}{MeV} threshold.

\subsubsection{Feature Importance Via Restricted-Feature Evaluation}
\label{sec:singleFeatureAnalysis}

The Deep Sets model used throughout this work is trained using all available point-cloud features (position, energy, and time). Since, unlike \glspl{bdt}, it does not provide a natural feature-importance ranking, an additional restricted-feature study is performed to estimate the contribution of each observable. Independent models are trained using topology only, topology plus timing, topology plus energy, and the complete feature set. This \textit{single-feature} evaluation isolates the contribution of each observable only approximately, since the cell time is computed as an energy-weighted average and is therefore intrinsically correlated with the deposited energy, even in the ``topology + time'' configuration.

The results for $p/\pi$ classification at \SI{10}{GeV} are reported in Table~\ref{tab:singleFeatures}. Topology alone already provides an accuracy of 74.59\%, showing that the shower shape carries significant discriminating information. Adding timing increases the accuracy to 80.45\%, whereas energy yields a much larger improvement to 91.46\%. Using all three observables further raises the accuracy to 93.80\%, indicating that energy provides most of the separation power, while timing contributes complementary information.

\begin{table}[ht]
    \centering
    \caption{%
    $p/\pi$ classification accuracy at \SI{10}{GeV} for different input feature sets assuming perfect timing resolution. The complete feature set is used throughout the main analysis, while the restricted feature sets are included to estimate the contribution of each observable.}
    \label{tab:singleFeatures}
    \vspace{4mm}
    \begin{tabular}{lcc}
        \hline
        \textbf{Features} & \textbf{68\% CL [\%]} & \textbf{Accuracy [\%]} \\
        \hline
        Topology                   & 74.39--74.90 & 74.59 \\
        Topology + Time            & 80.17--80.73 & 80.45 \\
        Topology + Energy          & 91.26--91.66 & 91.46 \\
        Topology + Energy + Time   & 93.63--93.97 & 93.80 \\
        \hline
    \end{tabular}
\end{table}

\subsubsection{Classification Performance As A Function Of Granularity And Energy}
\label{sec:deepsetsResults}

\begin{figure}[t]
    \centering
    \includegraphics[width=0.8\textwidth]{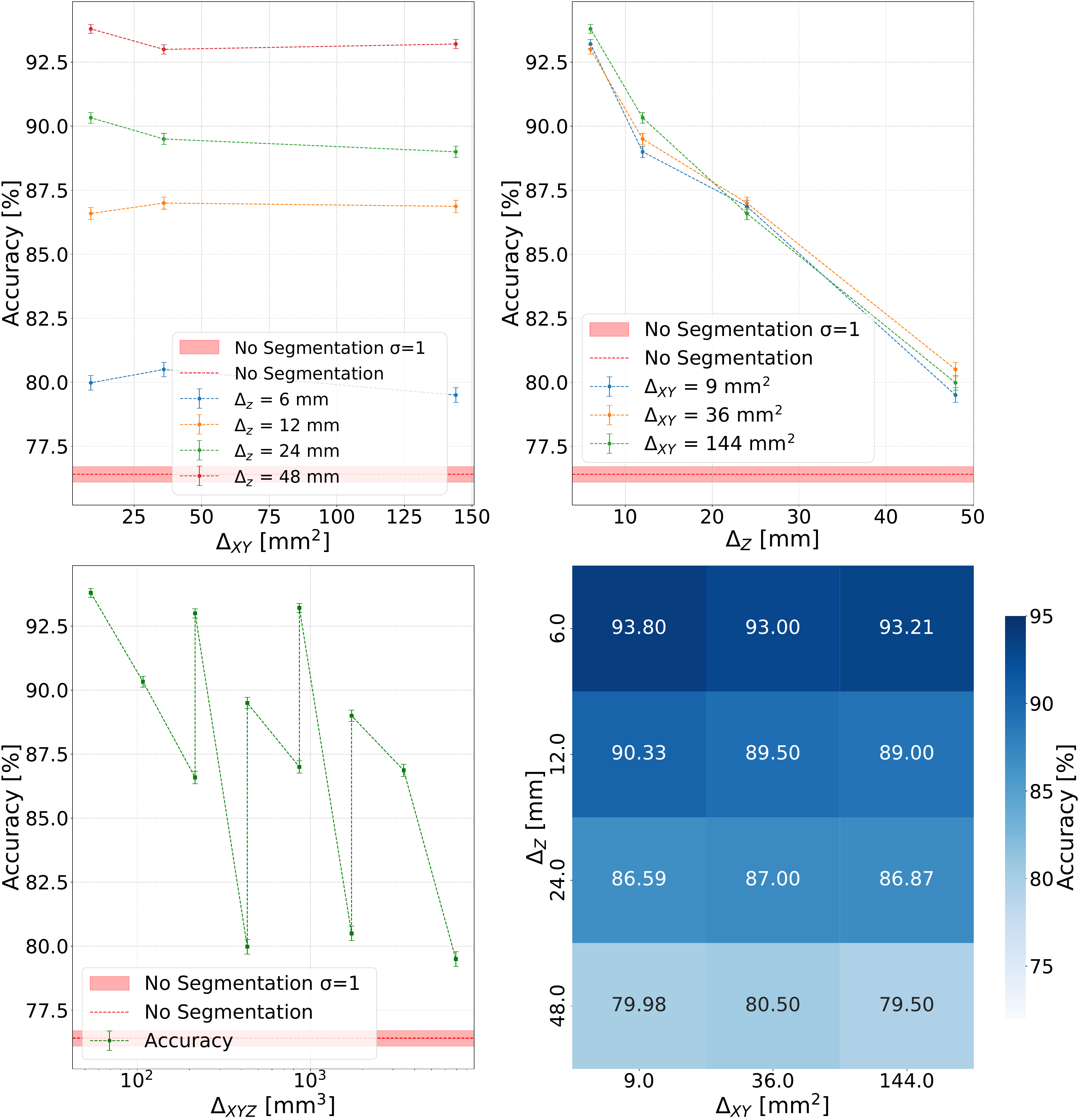}
    \caption{
Summary of $p/\pi$ classification results at \SI{10}{GeV}.
(\textbf{Top-Left}) Accuracy versus cell cross-section for different longitudinal segmentations.
(\textbf{Top-Right}) Accuracy versus cell length for different transverse segmentations.
(\textbf{Bottom-Left}) Accuracy versus cell volume.
The plotted $(\Delta_{xy},\Delta_z)$ pairs are
$\{(9,6),(9,12),(9,24),(9,48)\}$,
$\{(36,6),(36,12),(36,24),(36,48)\}$, and
$\{(144,6),(144,12),(144,24),(144,48)\}$,
in units of $(\si{\milli\metre\squared},\si{\milli\metre})$.
The corresponding values of $\Delta_{xyz}$, in
\si{\milli\metre\cubed}, are
$54$, $108$, $216$, $432$, $864$, $1728$, $3456$, and $6912$.
Configurations with equal volume but different aspect ratios produce the observed non-monotonic trend.
(\textbf{Bottom-Right}) Accuracy for different segmentation configurations.
Accuracy values are correlated because the same input events are shared across configurations.
}
    \label{fig:moneyplot_10GeV}
\end{figure}

\paragraph{Results At \SI{10}{GeV}.}
Figure~\ref{fig:moneyplot_10GeV} shows accuracy as a function of cell dimensions. Compared to the \gls{hlf} baseline at \SI{100}{GeV}, the Deep Sets model at \SI{10}{GeV} achieves substantially higher accuracy at all granularities. As expected, accuracy decreases monotonically with increasing cell volume (bottom left panel). Notably, performance is driven primarily by longitudinal segmentation: for a fixed $z$-cell size, varying the transverse segmentation produces only minor changes, while for a fixed transverse segmentation, reducing the $z$-cell size yields significant accuracy gains (top panels).

\begin{figure}[ht]
    \centering
    \includegraphics[
        height=0.24\textheight,
        keepaspectratio
    ]{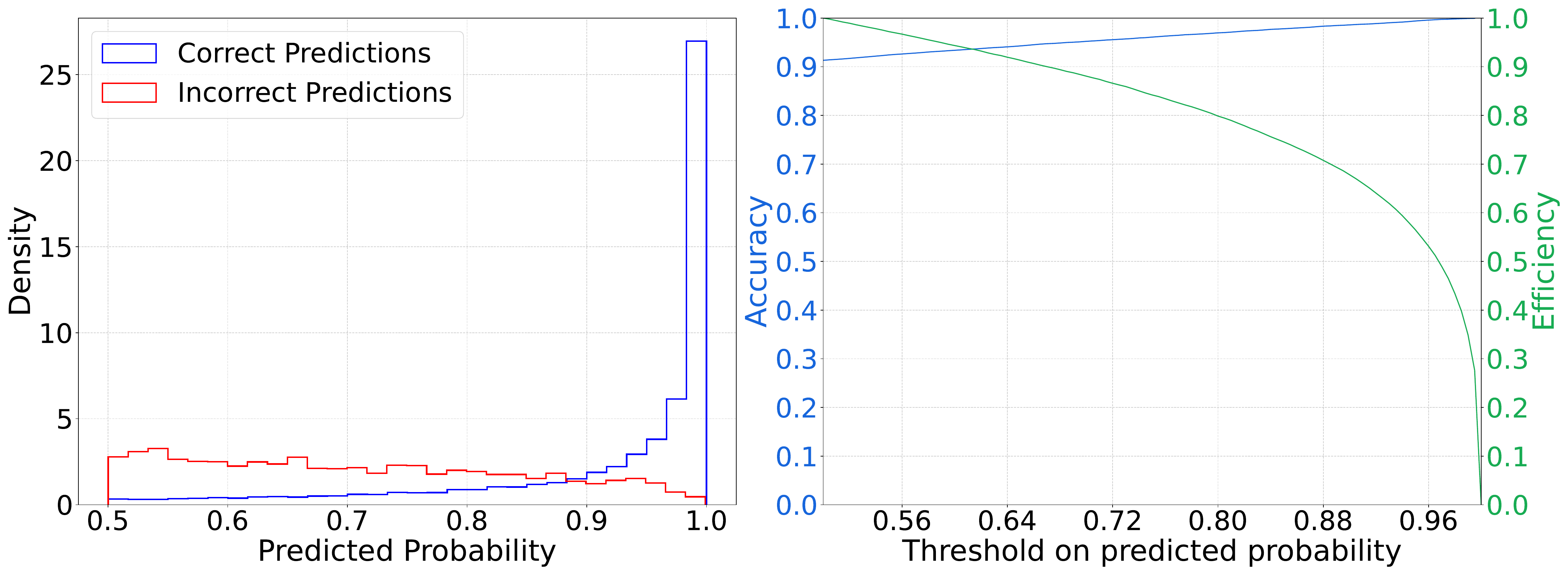}
    \caption{%
        Winning probability analysis for $p/\pi$ at \SI{10}{GeV} with segmentation $\{100,100,100\}$.
        \textbf{(Left)} Normalised distributions of the maximum predicted probability for correctly classified (blue) and misclassified (red) events.
        \textbf{(Right)} Accuracy and efficiency as a function of the probability threshold.
    }
    \label{fig:winningprobAndEfficiency10GeV}
\end{figure}

The reliability of the model output is assessed through the winning probability distributions (see Fig.~\ref{fig:winningprobAndEfficiency10GeV}). The distribution of correct predictions is clearly separated from that of incorrect ones, and a probability threshold can be defined to select a high-purity sample at the cost of reduced efficiency. This trade-off is relevant for applications such as \gls{pf} algorithms, where it may be preferable to discard ambiguous candidates rather than accept misidentified particles.

\FloatBarrier

\begin{figure}[ht]
    \centering
    \includegraphics[width=0.7\textwidth]{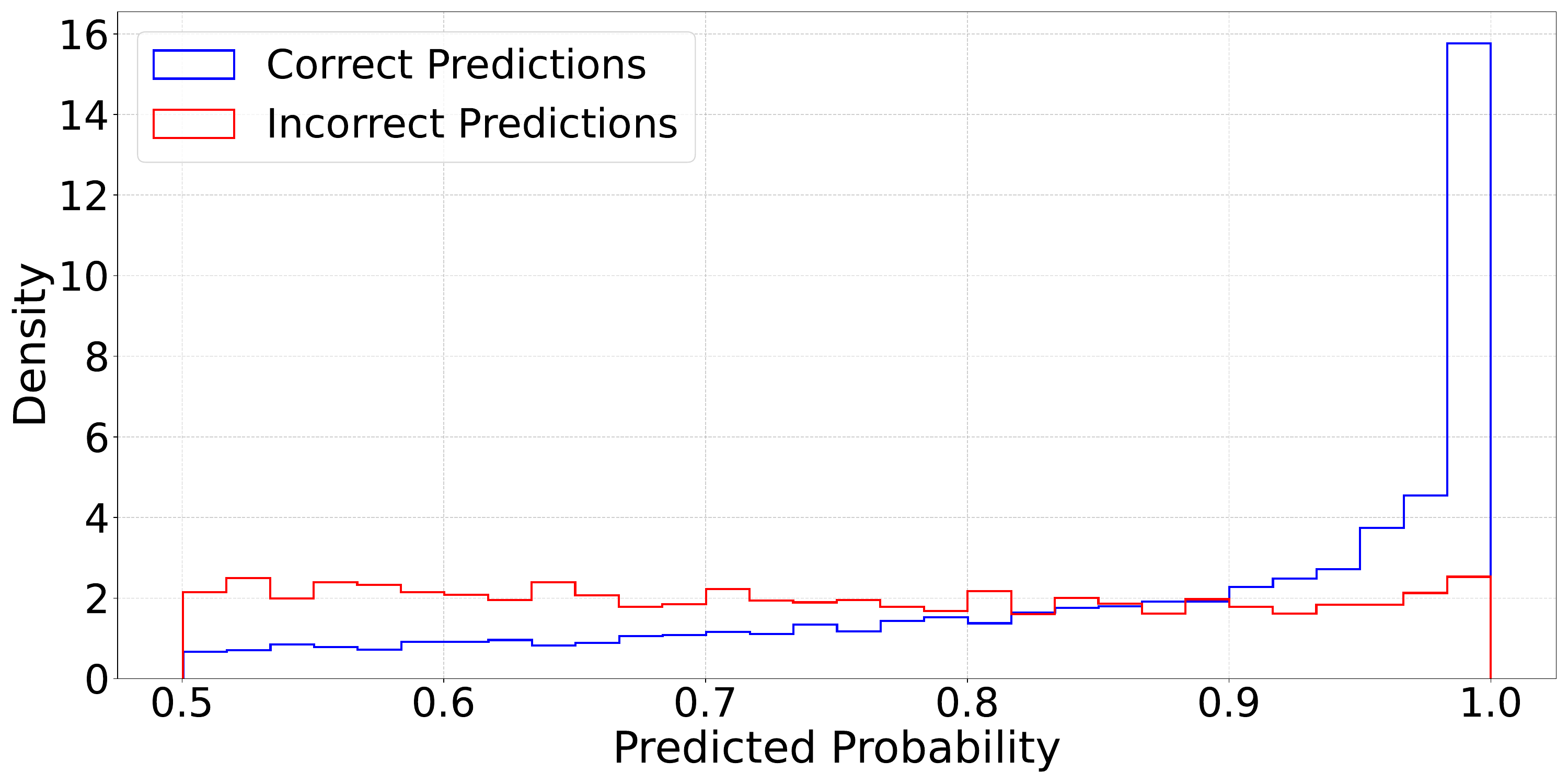}
    \caption{%
        Normalised winning probability distributions for the Deep Sets model at \SI{25}{GeV} with cell size $3\times3\times6\;\text{mm}^3$.
        Correctly classified events in blue, misclassified in red.
    }
    \label{fig:winning_prob_ppi_deepsets_25GeV}
\end{figure}

\begin{figure}[ht]
    \centering
    \includegraphics[width=0.8\textwidth]{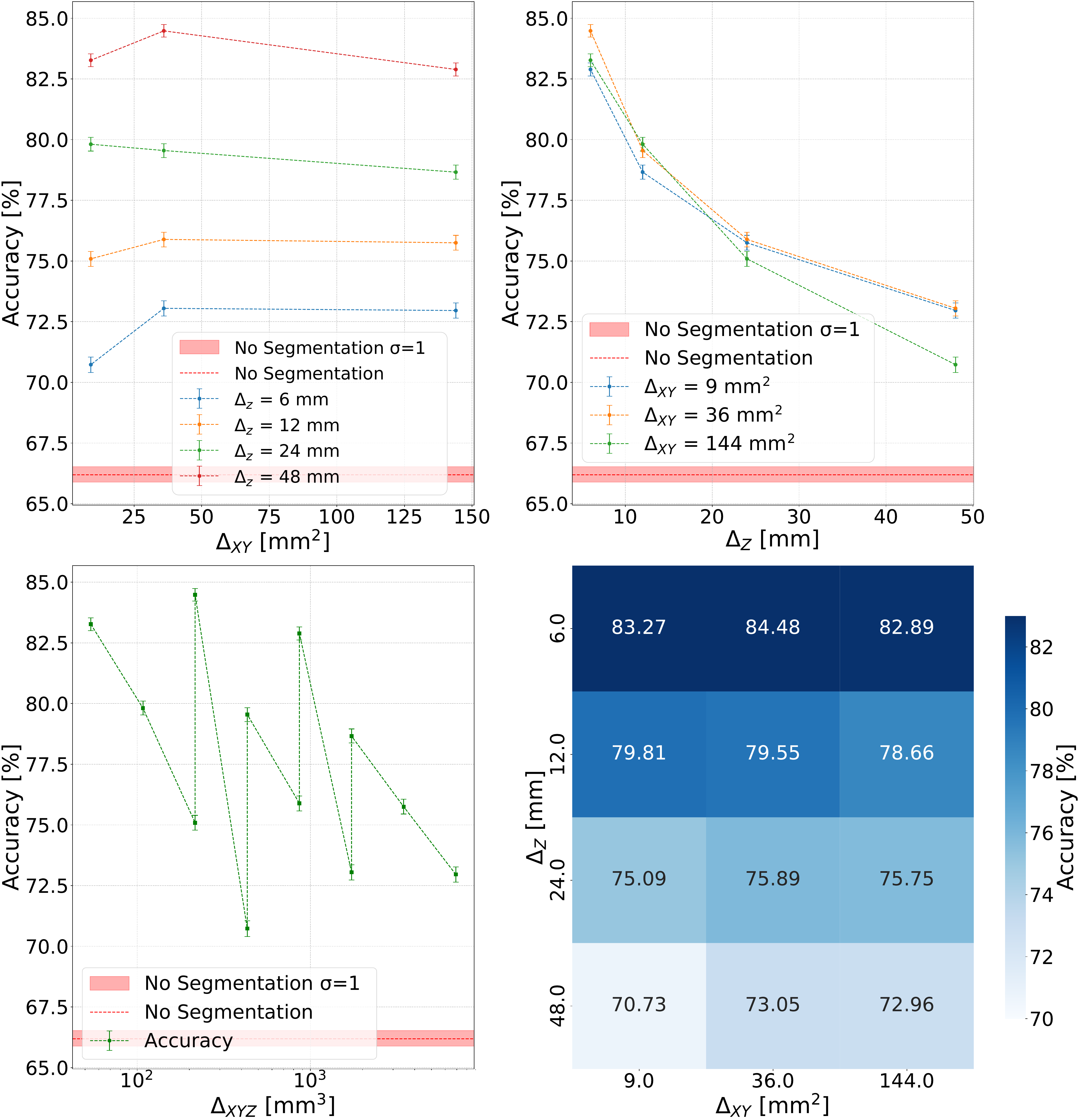}
    \caption{%
        Summary of $p/\pi$ classification results at \SI{25}{GeV}.
        Panel layout as in Fig.~\ref{fig:moneyplot_10GeV}.
        Accuracy values are correlated due to shared input data.
    }
    \label{fig:moneyplot_ppi_deepsets_25GeV}
\end{figure}

\paragraph{Results At \SI{25}{GeV}.}
At \SI{25}{GeV}, the winning probability distribution for correct predictions is strongly peaked near unity at the finest granularity (Fig.~\ref{fig:winning_prob_ppi_deepsets_25GeV}), confirming that the model operates in a high-confidence regime. The accuracy as a function of granularity (Figure~\ref{fig:moneyplot_ppi_deepsets_25GeV}) confirms the pattern observed at \SI{10}{GeV}: longitudinal segmentation is the dominant factor, while transverse segmentation contributes marginally.

\begin{figure}[ht]
    \centering
    \includegraphics[width=0.45\textwidth]{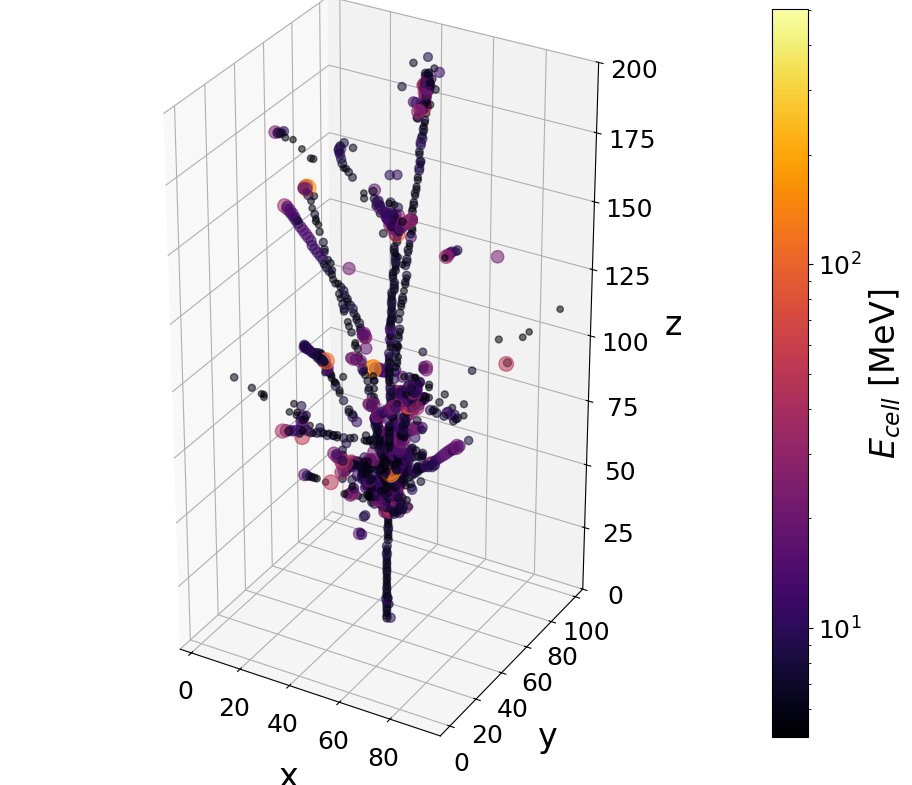}
    \hfill
    \includegraphics[width=0.45\textwidth]{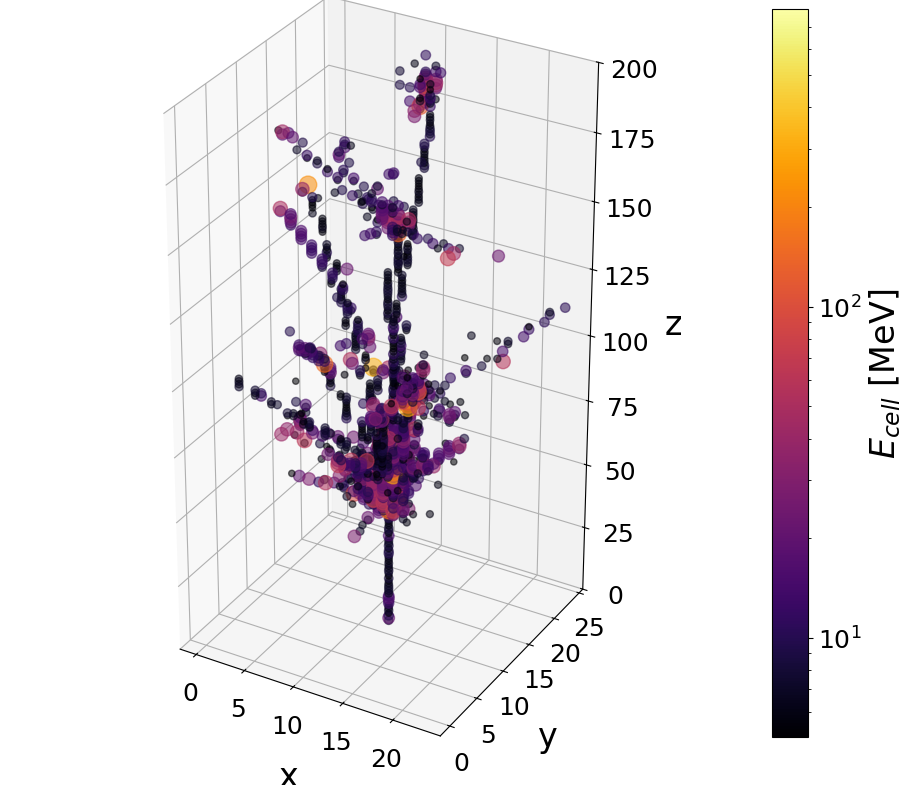}
    \caption{%
        Point-cloud representation of the same \SI{25}{GeV} proton shower
        at two transverse segmentations.
        \textbf{Left:} $\{100,100,200\}$.
        \textbf{Right:} $\{25,25,200\}$.
    }
    \label{fig:radialSegmentationEffect}
\end{figure}

The reduced sensitivity to transverse segmentation is physically motivated. Figure~\ref{fig:radialSegmentationEffect} shows that reducing the transverse granularity from $\{100,100\}$ to $\{25,25\}$ (in units of elementary cells) leaves the shower topology largely intact, as the most discriminating features arise from the longitudinal depth profile rather than the lateral spread. The energy threshold further reinforces this effect: by suppressing diffuse low-energy deposits, it preserves the topologically relevant structures even when coarser transverse segmentations are used.

\FloatBarrier

\begin{figure}[ht]
    \centering
    \includegraphics[width=0.8\textwidth]{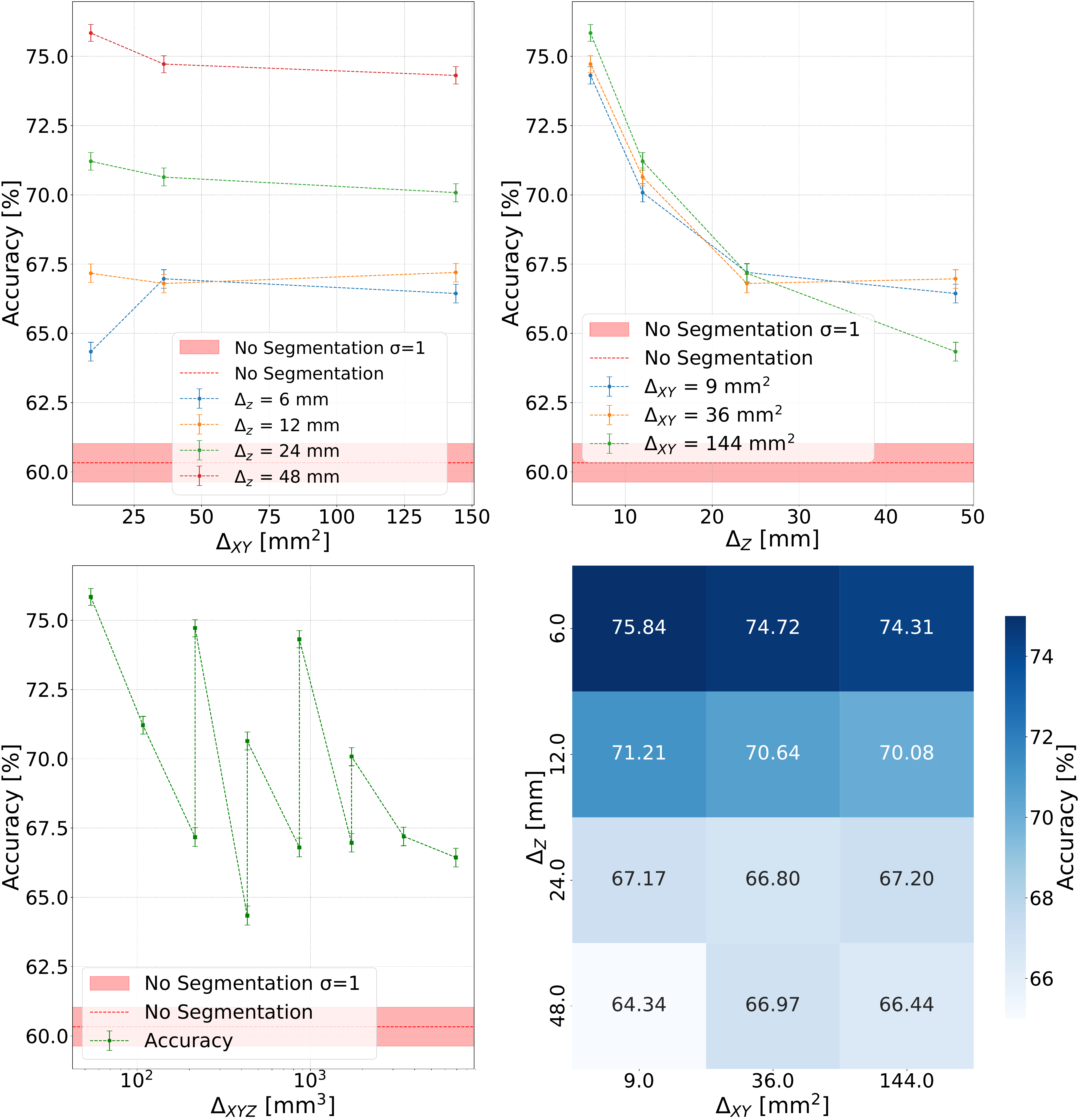}
    \caption{%
        Summary of $p/\pi$ classification results at \SI{50}{GeV}.
        Panel layout as in Fig.~\ref{fig:moneyplot_10GeV}.
        Accuracy values are correlated due to shared input data.
    }
    \label{fig:moneyplot_ppi_deepsets_50GeV}
\end{figure}

\begin{figure}[ht]
    \centering
    \includegraphics[width=0.7\textwidth]{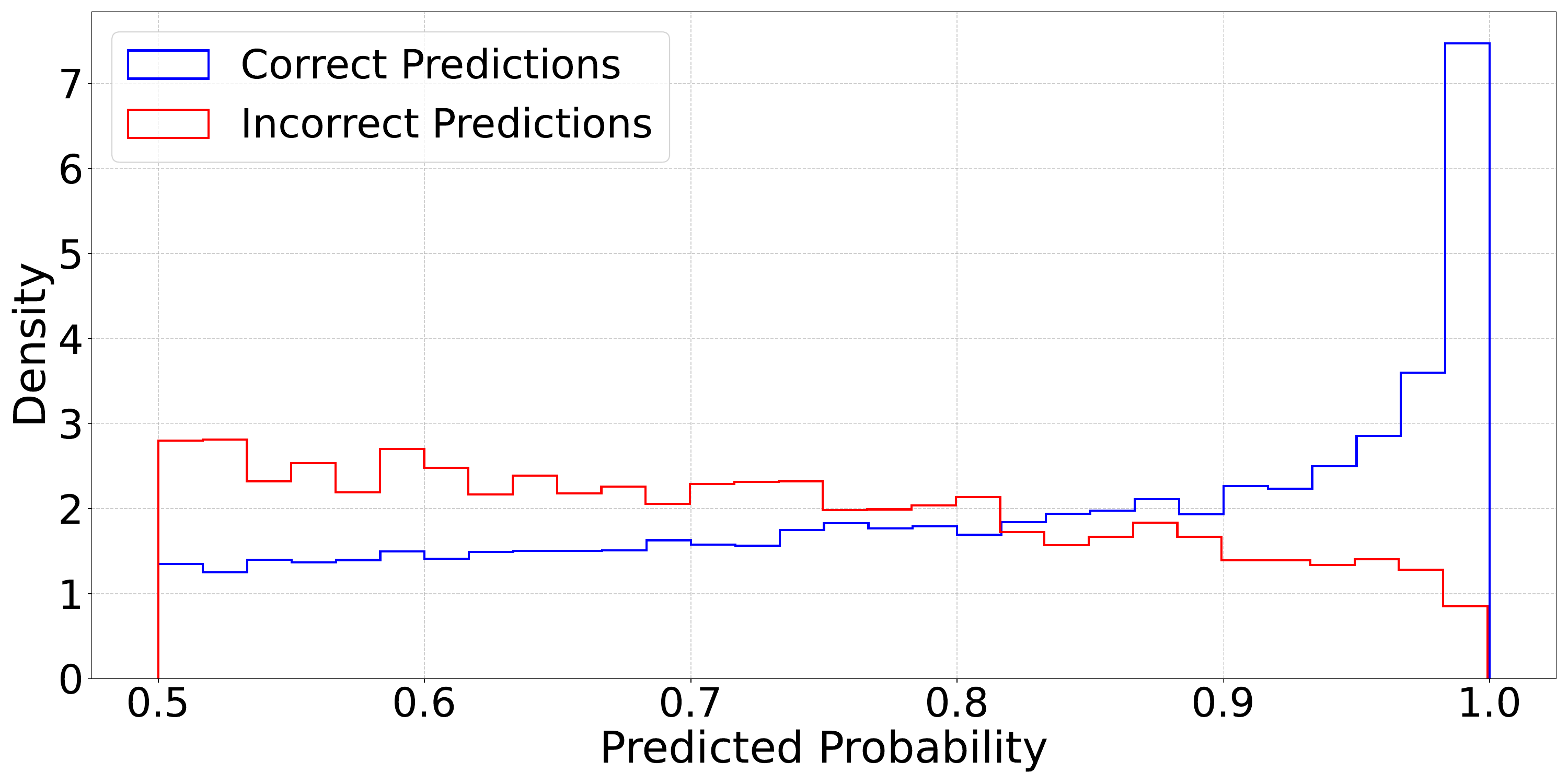}
    \caption{%
        Normalised winning probability distributions for the Deep Sets model at \SI{50}{GeV} with cell size $3\times3\times6\;\text{mm}^3$.
        Correctly classified events in blue, misclassified in red.
    }
    \label{fig:winning_prob_ppi_deepsets_50GeV}
\end{figure}

\paragraph{Results At \SI{50}{GeV}.}
Figure~\ref{fig:moneyplot_ppi_deepsets_50GeV} shows that the same qualitative picture holds at \SI{50}{GeV}: accuracy exceeds the baseline at all granularities, degrades with increasing cell volume, and is more sensitive to longitudinal than transverse segmentation. The winning probability distribution (Fig.~\ref{fig:winning_prob_ppi_deepsets_50GeV}) shows a correct-prediction peak near unity, though less sharp than at \SI{25}{GeV}. Consequently, applying a probability threshold yields a smaller absolute improvement in accuracy than at the lower energy point, reflecting the greater intrinsic difficulty of the classification task.
A practical implication follows from the relative importance of the two segmentation directions: for a fixed cell volume, prioritizing finer longitudinal segmentation over finer transverse segmentation is preferable, since the longitudinal shower profile provides more discriminating information for a given level of detector complexity.

\paragraph{Results At \SI{100}{GeV} And Comparison With The Baseline.}

\begin{figure}[ht]
    \centering
    \includegraphics[width=0.46\textwidth]{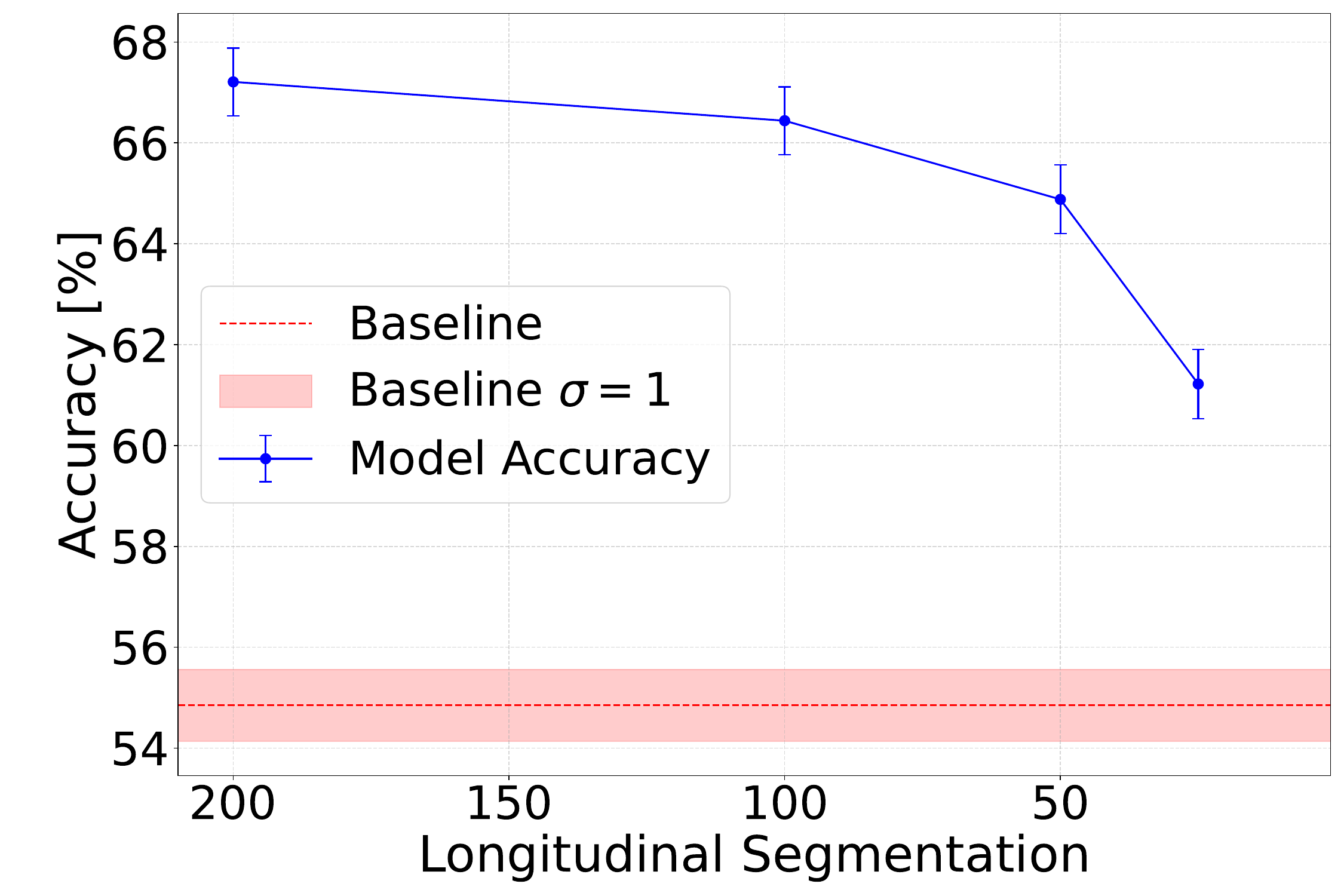}
    \hfill
    \includegraphics[width=0.46\textwidth]{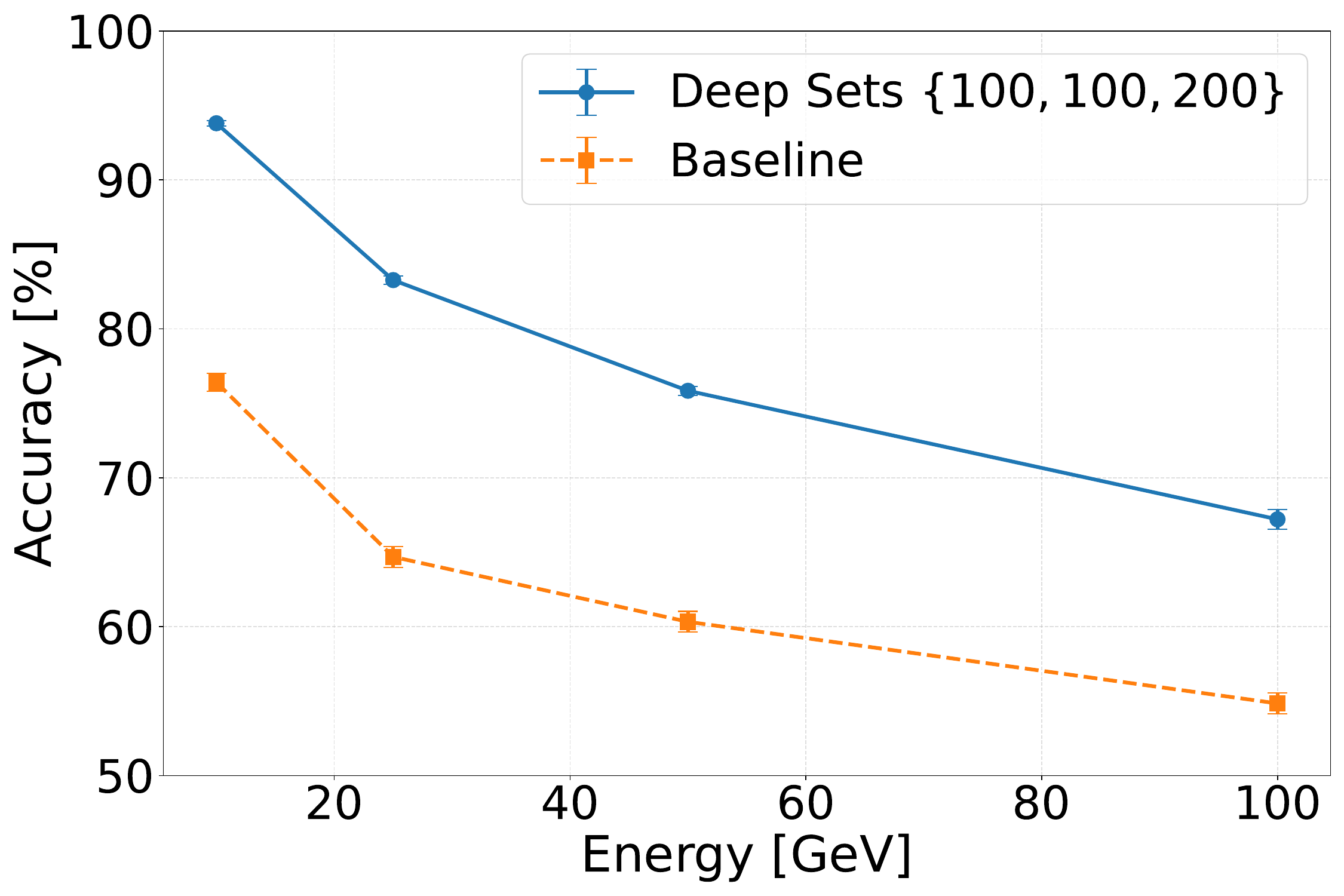}
    \caption{%
        \textbf{Left:} Accuracy of the Deep Sets model for $p/\pi$ at
        \SI{100}{GeV} as a function of longitudinal segmentation
        (transverse segmentation fixed at $\{100,100\}$). The \gls{bdt}
        result from Sec.~\ref{sec:resultsHLF} is shown for comparison.
        \textbf{Right:} Accuracy as a function of primary particle energy
        at maximum granularity $\{100,100,200\}$ (solid blue). The dashed
        orange line shows the baseline.
    }
    \label{fig:finalResults}
\end{figure}

Given the established dominance of the longitudinal dimension, the analysis at \SI{100}{GeV} fixes the transverse segmentation at $\{100,100\}$ and varies only the longitudinal segmentation (see Fig.~\ref{fig:finalResults} on the left). If compared with the results presented in Sec.~\ref{sec:resultsHLF}, Deep Sets achieves higher accuracy at all granularities, which can be attributed to two factors: it is trained on twice as many events, and the point cloud representation captures topological information that is inaccessible to aggregate \glspl{hlf}. It can be also noticed that the accuracy with "no-segmentation" using Deep Sets is lower than the accuracy using \gls{bdt} (see Fig.~\ref{fig:baselinePerformance} for comparison). This can be explained by the absence of the primary particle time-of-flight (\gls{tof}) in the point-cloud representation.

\paragraph{Evolution With Primary Particle Energy.}

The right plot of Fig.~\ref{fig:finalResults} summarizes the performance of the Deep Sets model at maximum granularity ($\{100,100,200\}$) as a function of the primary particle energy. Accuracy decreases monotonically from 93.8\% at \SI{10}{GeV} to 67.2\% at \SI{100}{GeV}.

This trend may be attributed to the increasing complexity of high-energy hadronic showers. It should not be interpreted as an increase in the interaction probability of the primary particle, since the nuclear interaction length is approximately energy independent over the energy range considered. Instead, higher-energy interactions generally produce more numerous and more energetic secondary particles, which are themselves more likely to undergo additional interactions and generate distinguishable sub-showers. Each additional vertex initiates a secondary sub-shower, and the observable shower properties, such as radius, depth profile, energy-weighted moments, become averages over these multiple contributions rather than properties of a single primary interaction. This averaging makes the shower characteristics of protons and pions more similar, reducing the discrimination power of any classifier. A more targeted analysis of shower substructure would be required to recover sensitivity at high energy, and such an analysis would itself require fine detector granularity. These results therefore provide an indirect demonstration that high-energy \gls{pid} would not be feasible without a sufficiently granular calorimeter.

\section{Discussion}
\label{sec:discussion}

The picture emerging from the studies and results discussed in the previous sections shows how highly granular calorimeters encode significantly richer information than is traditionally exploited in event reconstruction. The details of spatial development of hadronic showers contain measurable signatures of the identity of the incident particles. That information survives to the detection level through the different physical mechanisms that these particles and their daughters undergo in their interaction with the detector material, reflecting the different cross sections of the involved phenomena. Understanding how much discrimination is attainable in an instrument of characteristics exceeding the present \acrshort{sota} is an important step toward fine-tuning the design of future instruments such that they can be maximally sensitive to particle identity.

Modern machine-learning methods, such as those introduced in Sec.~\ref{sec:ml_strategy}, provide a means of exploiting the correlations among these shower properties. The two approaches considered in this work offer complementary perspectives. The \gls{bdt} analysis identifies physically interpretable differences through reconstructed \glspl{hlf}, whereas Deep Sets accesses the cell-level shower representation directly. The importance assigned by the \gls{bdt} to the total deposited energy, shower radius, and energy-weighted radial moments confirms that proton--pion discrimination is not driven by a single observable, but by correlated differences in the energy distribution and shower topology. The stronger performance of Deep Sets for segmented configurations further indicates that the cell-level representation retains information that is partially lost when a shower is compressed into a limited set of global observables. Nevertheless, the numerical difference between the two approaches should not be interpreted as a controlled comparison of their architectures, since their training samples and timing treatments are not identical.

A central result is the substantially stronger dependence of the classification performance on longitudinal segmentation than on transverse segmentation. Within the geometries considered, increasing the cell dimension along the shower direction produces a considerably larger loss in accuracy than a comparable coarsening of the transverse dimensions. This indicates that an important part of the proton--pion separation is encoded in the sequence of energy deposits and secondary interactions along the shower development. Consequently, cell volume alone is not sufficient to characterize the information retained by a detector: cells with similar volumes but different aspect ratios yield significantly different \gls{pid} performance. For isolated particles incident normally on the calorimeter, allocating segmentation preferentially along the longitudinal direction may therefore provide a more favorable balance between information content and channel count.

The above observation does not imply that transverse granularity is generally of secondary importance. The present study considers isolated particles entering the detector orthogonally, without pileup or nearby showers. In a realistic collider environment, transverse segmentation remains essential for resolving overlapping showers, associating calorimeter deposits with reconstructed tracks, and supporting \gls{pf} and jet-substructure algorithms. The relative importance of longitudinal and transverse segmentation must therefore be reassessed in realistic multi-particle events.

The potential applications of calorimeter-based \gls{pid} extend beyond the binary classification studied here. \gls{pf} reconstruction may benefit from knowledge of the species entering the calorimeter, for example through particle-dependent energy calibration or improved track--cluster association. The identification of energetic kaons within jets could contribute to strange-quark tagging, while a more detailed characterization of hadronic showers could assist boosted-object reconstruction, where several nearby particles produce overlapping showers, and the construction of particle-enriched calibration samples. These possibilities remain prospective and require dedicated studies in realistic experimental environments.

The results also have implications for detector optimization. Current calorimeter-design studies generally optimize segmentation with respect to energy resolution, reconstruction efficiency, shower separation, pattern recognition, and the associated detector cost. The present findings suggest that \gls{pid} sensitivity should be included as an additional objective. Because increasing segmentation entails substantial costs in channel count, readout complexity, data throughput, and computing resources, the relevant optimization is inherently multi-objective. Particle-identification performance could therefore be incorporated into the Pareto front or global utility function used to compare future detector concepts, together with the more traditional performance metrics.

One final consideration concerns the limitations of the present study. Although \texttt{GEANT4} simulations are generally regarded as high fidelity within the \gls{hep} community, some shower properties exploited by the discrimination algorithms may not perfectly reproduce the underlying physical processes. Moreover, the detector readout and electronics response were not modeled; including effects such as noise, thresholds, saturation, and finite measurement resolution would likely reduce the reported performance. The simulations also consider isolated single-particle showers with normal incidence and do not include pileup or overlapping showers, both of which would make the identification task considerably more challenging. Finally, to study the contributions of the different components of the five-dimensional per-deposit input \((E, t, x, y, z)\) to the Deep Sets classification, we assumed perfect time resolution. Since such resolution cannot be achieved by a physical detector, results involving timing information should be interpreted as an idealized upper bound. Nevertheless, although the classification performance obtained using topological information alone is significantly lower, it remains sufficient to support the qualitative conclusions discussed in this section.

The reported accuracies should therefore be regarded as optimistic benchmarks for the simulated configurations rather than as performance levels directly attainable in a realistic detector. At the same time, they are not strict upper bounds on the information available in the calorimeter, since alternative architectures may recover correlations not fully exploited by the models considered here. 

\section{Conclusions}
\label{sec:conclusions}

In this work we investigated how much information about the identity of an incident hadron can be extracted from the spatial, energetic, and temporal structure of its interactions in a highly granular homogeneous calorimeter. We also studied how the accessibility of that information depends on the dimensions of the independently read out cells. Our analysis focused on the discrimination of protons from positively charged pions, building on the broader\footnote {In Ref.~\cite{DeVita2025}, we also considered the identification of positive kaons, which is generally more challenging. Here we focus on protons versus pions in order to dig deeper in the above research questions.} particle-identification study presented in Ref.~\cite{DeVita2025}. These particles provide a well-defined benchmark because their distinct interactions with detector material produce measurable differences in hadronic-shower development.

The results establish that substantial proton--pion discrimination is possible from calorimeter information alone. At \SI{10}{GeV}, the Deep Sets model reaches an accuracy of 93.8\% for the finest segmentation considered, corresponding to cells of \(3 \times 3 \times 6\,\mathrm{mm}^3\). The restricted-feature study shows that shower topology is informative by itself, while the deposited energy provides the largest additional improvement. Timing supplies complementary information under the idealized assumptions adopted in the point-cloud analysis, although a scan incorporating realistic timing resolutions is required before drawing conclusions about the ultimate detector performance achievable in a real instrument.

Detector segmentation has a pronounced effect on the amount of \gls{pid} information that can be recovered. Coarse cells merge distinct parts of the shower and suppress correlations associated with its fine structure. This loss cannot be described solely in terms of cell volume: the cell dimensions along different directions play distinct roles. Classification performance is considerably more sensitive to the longitudinal size of cells than to their transverse dimensions, indicating that the evolution of the shower along its propagation direction contains a particularly important fraction of the discriminating information.

The achievable performance also depends strongly on the primary-particle energy. Even at maximum granularity, the classification accuracy decreases monotonically from 93.8\% at \SI{10}{GeV} to 67.2\% at \SI{100}{GeV}. The increasing complexity of higher-energy hadronic cascades makes the shower properties of protons and pions more similar and reduces their discriminability. Recovering additional sensitivity at high energy may therefore require methods that explicitly resolve secondary interaction vertices and local shower substructure.

These findings indicate that future calorimeters may provide \gls{pid} information complementary to that obtained from dedicated systems based on specific ionization, time of flight, or Cherenkov radiation. Determining how effectively this information can improve \gls{pf} reconstruction, particle-dependent calibration, and flavour-tagging algorithms remains an important direction for future work. As increasingly granular calorimeter technologies are developed, \gls{pid} sensitivity should be considered alongside energy resolution, shower separation, reconstruction efficiency, and cost in the co-design and optimization of future detector systems~\cite{MODE:2026xoy}.

\section*{Acknowledgements}

PV is supported by the “Ramón y Cajal” program under Project No. RYC2021-033305-I funded by the MCIN MCIN/AEI/10.13039/501100011033 and by the European Union NextGenerationEU/PRTR, and by the European Innovation Council (EIC) Pathfinder project PHINDER, grant agreement No. 101258353, funded by the European Union.

The authors gratefully acknowledge the computer resources at Artemisa and the technical support provided by the Instituto de Fisica Corpuscular, IFIC (CSIC-UV). Artemisa is co-funded by the European Union through the 2014-2020 ERDF Operative Programme of Comunitat Valenciana, project IDIFEDER/2018/048. 

\begingroup

\setlength{\LTleft}{0pt}
\setlength{\LTright}{\fill}

\addtolength{\glsdescwidth}{-2.5cm}
\addtolength{\glspagelistwidth}{2.5cm}

\printglossary[
    type=\acronymtype,
    title={Glossary},
    style=long3col
]

\endgroup

\markboth{Glossary}{}
\markright{}         

\bibliographystyle{JHEP}
\bibliography{biblio.bib}

@article{DeVita2025,
  author  = {De Vita, A. and Abhishek and Aehle, M. and Awais, M. and Breccia, A. and others},
  title   = {Hadron identification prospects with granular calorimeters},
  journal = {Particles},
  year    = {2025},
  volume  = {8},
  number  = {2},
  doi     = {10.3390/particles8020058},
  url     = {https://doi.org/10.3390/particles8020058}
}

@mastersthesis{DeVita2024Thesis,
  author  = {De Vita, A.},
  title   = {Hadron identification in highly granular calorimeters},
  school  = {University of Padova},
  year    = {2024},
  type    = {"Master's thesis"},
  address = {Padova, Italy},
  note   = {"Physics of Data, Department of Physics and Astronomy ``Galileo Galilei'', available at \url{https://thesis.unipd.it/handle/20.500.12608/91173}"}
  }

@inproceedings{Chen2016,
  author    = {Chen, T. and Guestrin, C.},
  title     = {{XGBoost}: A scalable tree boosting system},
  booktitle = {Proceedings of the 22nd {ACM SIGKDD} International Conference on Knowledge Discovery and Data Mining},
  year      = {2016},
  publisher = {Association for Computing Machinery},
  address   = {New York, NY, USA},
  series    = {KDD '16},
  isbn      = {9781450342322},
  doi       = {10.1145/2939672.2939785},
  url       = {https://doi.org/10.1145/2939672.2939785}
}

@article{Friedman2001,
  author  = {Friedman, J. H.},
  title   = {Greedy function approximation: A gradient boosting machine},
  journal = {The Annals of Statistics},
  year    = {2001},
  volume  = {29},
  doi     = {10.1214/aos/1013203451},
  url     = {https://doi.org/10.1214/aos/1013203451}
}

@misc{zaheer2018,
  author        = {Zaheer, M. and Kottur, S. and Ravanbakhsh, S. and Poczos, B. and Salakhutdinov, R. and Smola, A.},
  title         = {Deep sets},
  year          = {2018},
  eprint        = {1703.06114},
  archiveprefix = {arXiv},
  url           = {https://arxiv.org/abs/1703.06114}
}

@article{PerezLara2024,
  author  = {Perez-Lara, C. and Wetzel, J. and Akgun, U. and Anderson, T. and Barbera, T. and Blend, D. and Cankocak, K. and Cerci, S. and Chigurupati, N. and Cox, B. and Debbins, P. and Dubnowski, M. and Duran, B. and Dincer, G. G. and Hatipoglu, S.},
  title   = {Study of time resolution measurements and prospects for energy resolution of an ultra-compact sampling calorimeter ({RADiCAL}) module at {EM} shower maximum over the energy range 25--150 {GeV}},
  journal = {Nuclear Instruments and Methods in Physics Research Section A: Accelerators, Spectrometers, Detectors and Associated Equipment},
  year    = {2024},
  volume  = {1068},
  doi     = {10.1016/j.nima.2024.169737},
  url     = {https://doi.org/10.1016/j.nima.2024.169737}
}

@article{Lin1991,
  author  = {Lin, J.},
  title   = {Divergence measures based on the {Shannon} entropy},
  journal = {IEEE Transactions on Information Theory},
  year    = {1991},
  volume  = {37},
  number  = {1},
  doi     = {10.1109/18.61115},
  url     = {https://doi.org/10.1109/18.61115}
}

@article{RevModPhys.75.1243,
  author  = {Fabjan, C. W. and Gianotti, F.},
  title   = {Calorimetry for particle physics},
  journal = {Reviews of Modern Physics},
  year    = {2003},
  volume  = {75},
  number  = {4},
  doi     = {10.1103/RevModPhys.75.1243},
  url     = {https://doi.org/10.1103/RevModPhys.75.1243}
}

@article{Aad:1694142,
  author        = {{ATLAS Collaboration}},
  title         = {Electron reconstruction and identification efficiency measurements with the {ATLAS} detector using the 2011 {LHC} proton--proton collision data},
  journal       = {The European Physical Journal C},
  year          = {2014},
  volume        = {74},
  reportnumber  = {CERN-PH-EP-2014-040},
  eprint        = {1404.2240},
  archiveprefix = {arXiv},
  primaryclass  = {hep-ex},
  doi           = {10.1140/epjc/s10052-014-2941-0},
  url           = {https://doi.org/10.1140/epjc/s10052-014-2941-0}
}

@article{aad2017topocluster,
  author        = {{ATLAS Collaboration}},
  title         = {Topological cell clustering in the {ATLAS} calorimeters and its performance in {LHC} Run 1},
  journal       = {The European Physical Journal C},
  year          = {2017},
  volume        = {77},
  eprint        = {1603.02934},
  archiveprefix = {arXiv},
  primaryclass  = {hep-ex},
  doi           = {10.1140/epjc/s10052-017-5004-5},
  url           = {https://doi.org/10.1140/epjc/s10052-017-5004-5}
}

@article{Adloff2012,
  author  = {Adloff, C. and others},
  title   = {Hadronic energy resolution of a highly granular scintillator--steel hadron calorimeter using software compensation techniques},
  journal = {Journal of Instrumentation},
  year    = {2012},
  volume  = {7},
  number  = {9},
  doi     = {10.1088/1748-0221/7/09/P09017},
  url     = {https://doi.org/10.1088/1748-0221/7/09/P09017}
}

@article{Belayneh_2020,
  author  = {Belayneh, D. and others},
  title   = {Calorimetry with deep learning: Particle simulation and reconstruction for collider physics},
  journal = {The European Physical Journal C},
  year    = {2020},
  volume  = {80},
  number  = {7},
  doi     = {10.1140/epjc/s10052-020-8251-9},
  url     = {https://doi.org/10.1140/epjc/s10052-020-8251-9}
}

@article{qasim2022multiparticle,
  author  = {Qasim, S. R. and Chernyavskaya, N. and Kieseler, J. and Long, K. and Viazlo, O. and Pierini, M. and Nawaz, R.},
  title   = {End-to-end multi-particle reconstruction in high-occupancy imaging calorimeters with graph neural networks},
  journal = {The European Physical Journal C},
  year    = {2022},
  volume  = {82},
  number  = {8},
  doi     = {10.1140/epjc/s10052-022-10665-7},
  url     = {https://doi.org/10.1140/epjc/s10052-022-10665-7}
}

@article{kieseler2020object,
  author  = {Kieseler, J.},
  title   = {Object condensation: One-stage grid-free multi-object reconstruction in physics detectors, graph, and image data},
  journal = {The European Physical Journal C},
  year    = {2020},
  volume  = {80},
  number  = {9},
  doi     = {10.1140/epjc/s10052-020-08461-2},
  url     = {https://doi.org/10.1140/epjc/s10052-020-08461-2}
}

@article{Akchurin1998,
  author  = {Akchurin, N. and Ayan, S. and Bencze, G. L. and Chikin, K. and Cohn, H. and Doulas, S. and Dumanoǧlu, I. and Eskut, E. and Fenyvesi, A. and Ferrando, A.},
  title   = {On the differences between high-energy proton and pion showers and their signals in a non-compensating calorimeter},
  journal = {Nuclear Instruments and Methods in Physics Research Section A: Accelerators, Spectrometers, Detectors and Associated Equipment},
  year    = {1998},
  volume  = {408},
  number  = {2--3},
  doi     = {10.1016/S0168-9002(98)00021-7},
  url     = {https://doi.org/10.1016/S0168-9002(98)00021-7}
}

@misc{krause_2026_21626667,
  author    = {Krause, C. and Winterhalder, R. and Feickert, M. and Nachman, B. and Raine, J. and {HEP ML Community}},
  title     = {A living review of machine learning for particle physics},
  year      = {2026},
  month     = jul,
  publisher = {Zenodo},
  version   = {v2026.06.01},
  doi       = {10.5281/zenodo.21626667},
  url       = {https://doi.org/10.5281/zenodo.21626667}
}

@article{MODE:2026xoy,
    author = "Dorigo, Tommaso and Vischia, Pietro and others",
    title = "{On the Codesign of Scientific Experiments and Industrial Systems}",
    eprint = "2603.26613",
    archivePrefix = "arXiv",
    primaryClass = "physics.ins-det",
    month = "3",
    year = "2026"
}

@article{doi:10.1142/S0217751X26300115,
author = {Takahashi, F. and others},
collaboration = {Particle Data Group},
title = {Review of Particle Physics},
journal = {International Journal of Modern Physics A},
volume = {41},
number = {22},
pages = {2630011},
year = {2026},
doi = {10.1142/S0217751X26300115},
URL = {https://doi.org/10.1142/S0217751X26300115},
eprint = {https://doi.org/10.1142/0217751X26300115},
}

@article{KLEMPT1999542,
title = {Review of particle identification by time of flight techniques},
journal = {NIM A},
volume = {433},
number = {1},
pages = {542-553},
year = {1999},
issn = {0168-9002},
doi = {https://doi.org/10.1016/S0168-9002(99)00323-X},
url = {https://www.sciencedirect.com/science/article/pii/S016890029900323X},
author = {W. Klempt},
}

@article{PAPANESTIS2020162004,
author = {A. Papanestis},
title = {Cherenkov light imaging in particle and nuclear physics experiments},
journal = {NIM A},
volume = {952},
pages = {162004},
year = {2020},
issn = {0168-9002},
doi = {https://doi.org/10.1016/j.nima.2019.03.059},
url = {https://www.sciencedirect.com/science/article/pii/S0168900219303845},
}

@article{ANDRONIC2012130,
author = {A. Andronic and J.P. Wessels},
title = {Transition radiation detectors},
journal = {NIM A},
volume = {666},
pages = {130-147},
year = {2012},
issn = {0168-9002},
doi = {https://doi.org/10.1016/j.nima.2011.09.041},
url = {https://www.sciencedirect.com/science/article/pii/S0168900211018134},
}

@article{mueloss1,
  author  = {Kieseler, J. and Strong, G. C. and Chiandotto, F. and Dorigo, T. and Layer, L.},
  title   = {Calorimetric measurement of multi-{TeV} muons via deep regression},
  journal = {The European Physical Journal C},
  year    = {2022},
  volume  = {82},
  number  = {1},
  pages   = {79},
  doi     = {10.1140/epjc/s10052-022-09993-5},
  url     = {https://doi.org/10.1140/epjc/s10052-022-09993-5}
}

@misc{mueloss2,
  author        = {Dorigo, T. and Guglielmini, S. and Kieseler, J. and Layer, L. and Strong, G. C.},
  title         = {Deep regression of muon energy with a {K}-nearest neighbor algorithm},
  year          = {2022},
  eprint        = {2203.02841},
  archiveprefix = {arXiv},
  primaryclass  = {hep-ex},
  doi           = {10.48550/arXiv.2203.02841},
  url           = {https://arxiv.org/abs/2203.02841}
}

\end{document}